\documentclass[preprint2,twocolappendix,trackchanges]{aastex6}

\usepackage{amsmath,multirow,rotating}
\usepackage[graphicx]{realboxes}
\hypersetup{breaklinks=true,colorlinks=true,citecolor=blue,linkcolor=blue,urlcolor=blue}

\shorttitle{SN Rates in LOSS, I}
\shortauthors{Graur et al.}

\begin{document}

\title{LOSS Revisited --- I: Unraveling Correlations between Supernova Rates and Galaxy Properties, as Measured in a Reanalysis of the Lick Observatory Supernova Search}

\author{Or Graur\altaffilmark{1}}
\affil{Harvard-Smithsonian Center for Astrophysics, 60 Garden St., Cambridge, MA 02138, USA; \url{or.graur@cfa.harvard.edu}}
\affil{CCPP, New York University, 4 Washington Place, New York, NY 10003, USA}
\affil{Department of Astrophysics, American Museum of Natural History, New York, NY 10024, USA}

\author{Federica B. Bianco}
\affil{CCPP, New York University, 4 Washington Place, New York, NY 10003, USA}
\affil{Center for Urban Science and Progress, New York University, 1 MetroTech Center, Brooklyn, NY 11201, USA}

\author{Shan Huang and Maryam Modjaz}
\affil{CCPP, New York University, 4 Washington Place, New York, NY 10003, USA}

\author{Isaac Shivvers, Alexei V. Filippenko, and Weidong Li\altaffilmark{2}}
\affil{Department of Astronomy, University of California, Berkeley, CA 94720-3411, USA}

\author{J.~J. Eldridge}
\affil{Department of Physics, University of Auckland, Private Bag 92019, Auckland, New Zealand}

\altaffiltext{1}{NSF Astronomy and Astrophysics Postdoctoral Fellow.}
\altaffiltext{2}{Deceased 2011 December 12.}


\begin{abstract}
\noindent Most types of supernovae (SNe) have yet to be connected with their progenitor stellar systems. Here, we reanalyze the 10-year SN sample collected during 1998--2008 by the Lick Observatory Supernova Search (LOSS) in order to constrain the progenitors of SNe~Ia and stripped-envelope SNe (SE~SNe, i.e., SNe~IIb, Ib, Ic, and broad-lined Ic). We matched the LOSS galaxy sample with spectroscopy from the Sloan Digital Sky Survey and measured SN rates as a function of galaxy stellar mass, specific star formation rate, and oxygen abundance (metallicity). We find significant correlations between the SN rates and all three galaxy properties. The SN~Ia correlations are consistent with other measurements, as well as with our previous explanation of these measurements in the form of a combination of the SN~Ia delay-time distribution and the correlation between galaxy mass and age. The ratio between the SE~SN and SN~II rates declines significantly in low-mass galaxies. This rules out single stars as SE~SN progenitors, and is consistent with predictions from binary-system progenitor models. Using well-known galaxy scaling relations, any correlation between the rates and one of the galaxy properties examined here can be expressed as a correlation with the other two. These redundant correlations preclude us from establishing causality---that is, from ascertaining which of the galaxy properties (or their combination) is the physical driver for the difference between the SE~SN and SN~II rates. We outline several methods that have the potential to overcome this problem in future works.

\end{abstract}



\section{Introduction}
\label{sec:intro}

Of the various types of supernovae (SNe) we observe, only certain subtypes of SNe II have been conclusively connected to their progenitors. Pre-explosion imaging has shown that SNe II-plateau (SNe~IIP), for example, come from red supergiants (see review by \citealt{Smartt2009review}). Owing to their cosmological significance, the progenitor systems of SNe Ia have been pursued relentlessly over the last two decades, but their nature is still debated (see \citealt{2014ARA&A..52..107M} for a recent review).

SNe II and stripped-envelope SNe (SE SNe, i.e., SNe~IIb, Ib, Ic, and broad-lined Ic; e.g., \citealt{1997ARA&A..35..309F,2001AJ....121.1648M,2014AJ....147...99M}) are attributed to the core collapse (CC) of stars more massive than $\sim 8~{\rm M}_\sun$. Spectroscopically, SE SNe are distinguished from SNe II by the lack of hydrogen features (either partial, as in SNe IIb, or nearly complete, as in SNe Ib; e.g., \citealt{1997ARA&A..35..309F,2016ApJ...827...90L}). SNe Ic also lack helium features. In order to explain this lack of hydrogen and helium, SE~SNe are thought to be the explosions of stars that have had their outer envelopes stripped away before the explosion (hence their name). Of all the processes suggested to explain this stripping, the leading models make use of either stellar winds (e.g., \citealt{2003ApJ...591..288H}), interaction with a binary companion (e.g., \citealt{1971ARA&A...9..183P,1992ApJ...391..246P,1998A&A...333..557D,1998A&ARv...9...63V}), or a combination of both (e.g., \citealt{2011MNRAS.412.1522S}). In the case of broad-lined SNe Ic connected to gamma-ray bursts, chemically homogeneous evolution (e.g., \citealt{2005A&A...443..643Y,2006ApJ...637..914W}) and explosive common-envelope ejection \citep{2010MNRAS.406..840P} have also been suggested. Pre-explosion imaging of the sites of these SNe has so far failed to reveal the nature of their progenitors conclusively (e.g., \citealt{2013MNRAS.436..774E,2016MNRAS.461L.117E,2016ApJ...825L..22F,2016ApJ...818...75V}), though the case for yellow supergiants in binary systems as the progenitors of SNe IIb is gaining traction (e.g., \citealt{2013ApJ...772L..32V,2014AJ....147...37V,2014AJ....148...68B,2014A&A...565A.114F,2015MNRAS.446.2689E}). New observational methods are required to address this question.

SNe Ia are thought to be thermonuclear explosions of carbon--oxygen white-dwarf remnants of $<8$~${\rm M}_\sun$ stars. In order to disrupt the otherwise stable white dwarf, most models place it in a binary system where it can grow in mass, raising the temperature in the core until the carbon is ignited in a thermonuclear runaway. To grow in mass, the white dwarf can either siphon matter off of a main-sequence or evolved companion star (the so-called ``single-degenerate'' scenario; \citealt{Whelan1973}) or merge with a second carbon--oxygen white dwarf after the two spiral in owing to loss of energy and angular momentum to gravitational waves (the ``double degenerate'' scenario; \citealt{Iben1984,Webbink1984}). Recently, direct collisions of white dwarfs have also been suggested as a possible progenitor channel (e.g., \citealt{2012arXiv1211.4584K,2013ApJ...778L..37K,2015MNRAS.454L..61D}; but see also \citealt{2013MNRAS.430.2262H}).

Many methods have been used to constrain these various SN progenitor models, including (but far from limited to) direct imaging of the explosion sites either before or long after the SN explosion (e.g., \citealt{2008MNRAS.388..421M,Li2011fe,2014MNRAS.442L..28G,2016ApJ...819...31G,2014ApJ...790....3K}), multiwavelength follow-up observations (e.g., \citealt{2012ApJ...746...21H,2013ApJ...767...71M,2014ApJ...790...52M,2016ApJ...821..119C}), and analyses of SN remnants (e.g., \citealt{2004Natur.431.1069R,Schaefer2012,2013ApJ...774...99K,2014MNRAS.439..354B}).

Over the last few decades, studies have consistently shown that SNe Ia are more common in blue, star-forming, late-type galaxies than in red, passive, early-type galaxies (e.g., \citealt{1979AJ.....84..985O,1988A&A...190...10C,1989ApJ...345..752E,1990PASP..102.1318V,1991ARA&A..29..363V,1994ApJ...423L..31D,1997ApJ...483L..29W,cappellaro1999,2005ApJ...629..750D}). 

\citet{2006ApJ...648..868S} showed that SN Ia rates per unit mass decreased with increasing galaxy stellar mass. \citet[hereafter L11]{li2011rates} showed the same effect for all SN types, in all types of galaxies (but see Section~\ref{sec:SNIa}), and dubbed this the ``rate-size,'' or rate--mass, relation. We confirmed this trend for SNe~Ia in star-forming galaxies in \citet{GraurMaoz2013} and for SNe~II in \citet[hereafter G15]{2015MNRAS.450..905G}. Following \citet{Kistler2011}, we argued that the dependence of the SN Ia rates on stellar mass results from a combination of galaxy scaling relations (older galaxies are more massive, on average, than younger ones; \citealt{2005MNRAS.362...41G}) and the SN Ia delay-time distribution (DTD), which behaves as a power law with an index of $\sim -1$ (e.g., observations by \citealt{2008PASJ...60.1327T,Maoz2010clusters,Maoz2010magellan,Graur2011,Graur2014,2014AJ....148...13R}; and reviews by \citealt{2012NewAR..56..122W,2013FrPhy...8..116H,2014ARA&A..52..107M}). 

\citet{2005A&A...433..807M}, \citet{2006ApJ...648..868S}, \citet{2012ApJ...755...61S}, and G15 also measured SN Ia rates as a function of the galaxies' specific star-formation rate (sSFR). In G15, we showed that our explanation for the rate--mass correlation also explained the observed trend between SN Ia rates and sSFR, where the rates are constant in passive galaxies but rise with increasing sSFR in star-forming galaxies. 

In G15, we also measured SN rates as a function of stellar mass and sSFR for SNe II and claimed that their rate--mass relation was simply the result of their progenitors' short lifetimes: because SNe II come from massive ($>8~{\rm M}_\sun$) stars, their rates track the star-formation rates of their galaxies. Similar measurements of CC SN rates (i.e., combining SNe~II and SE~SNe) were made by \citet{botticella2012}.

L11 was part of a series of papers that explored the SN sample collected by the Lick Observatory Supernova Search (LOSS; L11; \citealt{2011MNRAS.412.1419L,li2011LF,Maoz2010loss,2011MNRAS.412.1522S}). LOSS is an ongoing survey for SNe in local galaxies using the 0.76 m Katzman Automatic Imaging Telescope. For detailed descriptions of the survey, see \citet{2000AIPC..522..103L}, \citet{2001ASPC..246..121F}, and \citet{2003fthp.conf..171F,2005ASPC..332...33F}. 

Here, we use the LOSS sample to remeasure and reanalyze the SN rates originally published by L11. In Section~\ref{sec:galaxies}, we match between the LOSS sample and spectroscopy from the Sloan Digital Sky Survey (SDSS; \citealt{2000AJ....120.1579Y}) in order to go beyond L11 and measure the SN rates not only as a function of galaxy stellar mass but also of sSFR and metallicity, as expressed by the abundance of gas-phase oxygen in the centers of the LOSS galaxies. Throughout this work, we use the metallicity scale of \citet[hereafter T04]{2004ApJ...613..898T}.

In Section~\ref{sec:addendum}, we make several addenda to the LOSS sample. We publish the control times necessary to measure SN rates with this sample, as well as updated tables of galaxy properties and SN rates. We deal with SNe Ia in Section~\ref{sec:SNIa}, and with CC SNe in Section~\ref{sec:CCSN}. 

We find significant correlations between the SN rates and the various galaxy properties. Most importantly, we find that the CC SN rates behave differently in different types of galaxies: the SE SN rates are shown to be depressed, relative to the SN II rates, in galaxies with low stellar mass, high sSFRs, and low metallicity values. Other studies have reported similar trends through measurements of correlations between fractions of SNe within a given sample and metallicity (e.g., \citealt{2008ApJ...673..999P,2009A&A...503..137B,2012ApJ...759..107K,2015PASA...32...19A}), or by splitting SN fractions between different types of galaxies, which encompass different metallicity regimes \citep{2010ApJ...721..777A,2014MNRAS.444.2428H}. We conduct an in-depth comparison of our results with these works in Section~\ref{sec:discuss}. We also show that our measurements rule out theoretical models based on single-star progenitors for SE SNe, but are consistent with models that assume binary-system progenitors. 

In Section~\ref{sec:discuss}, we additionally discuss how the various rate correlations are not independent. For all SN types, we show that a correlation with one galaxy property can be transformed into the measured correlations with any other of the galaxy properties studied here by using well-known galaxy scaling relations. This makes it impossible to distinguish causation from correlation, especially for the deficiency of SE SNe in lower-mass galaxies. However, we argue that the structure seen in the correlations (e.g., the way the ratio between the SE SN and SN II rates depends on galaxy stellar mass) can be incorporated into models and used to constrain progenitor models. 

We summarize our results in Section~\ref{sec:summary}. Paper II in this series \citep{2016arXiv160902923G} will use population fractions, as measured from the LOSS volume-limited subsample of SNe, to strengthen the results presented here and add further constraints on SN progenitor scenarios. For Paper II, we rely on a reclassification of the SNe in this subsample, as reported by \citet{2016arXiv160902922S}.


\section{Galaxy and Supernova Samples}
\label{sec:galaxies}

Between 1998 March and 2008 December, LOSS discovered a total of 1036 SNe. Most of these were discovered among the 14,882 galaxies directly targeted by the search (at a median distance of $80^{+50}_{-40}$ Mpc, where the upper and lower bounds contain 68\% of the galaxies in the sample), but a few dozen were also discovered in background galaxies in the LOSS fields. This sample, along with the subsamples used to measure the LOSS SN luminosity functions \citep{li2011LF} and rates (L11), is described in detail by \citet{2011MNRAS.412.1419L}. 

LOSS classified their SNe into three broad categories (e.g., \citealt{1997ARA&A..35..309F}): SNe Ia, Ib/c, and II. The first category included all SN~Ia subtypes, including the subluminous SN 1991bg-like SNe~Ia, overluminous SN 1991T-like SNe~Ia, and SN 2002cx-like SNe~Ia (now referred to as SNe~Iax; \citealt{2006AJ....132..189J,2013ApJ...767...57F}). SNe~Ib/c included SNe~Ib, Ic, and ``peculiar'' SE~SNe (such as broad-lined SNe~Ic; see, e.g., \citealt{2006ARA&A..44..507W} for a review). The final category comprised SNe~IIP, IIL, IIn, and IIb. However, as mentioned earlier, SNe~IIb are characterized by hydrogen deficiency, which is an indication of envelope stripping. Thus, they should be grouped with the SNe~Ib/c (e.g., \citealt{1993ApJ...415L.103F}). In this work, we use ``SE~SNe'' instead of ``SNe~Ib/c.'' However, because we use the control times calculated by L11 for the LOSS sample (see Sections~\ref{sec:addendum} and \ref{subsec:rates_mass}, below), we must keep the SNe~IIb grouped with the SN~II class when calculating rates. 

\begin{figure*}
 \center
 \begin{tabular}{cc}
  \includegraphics[width=0.47\textwidth]{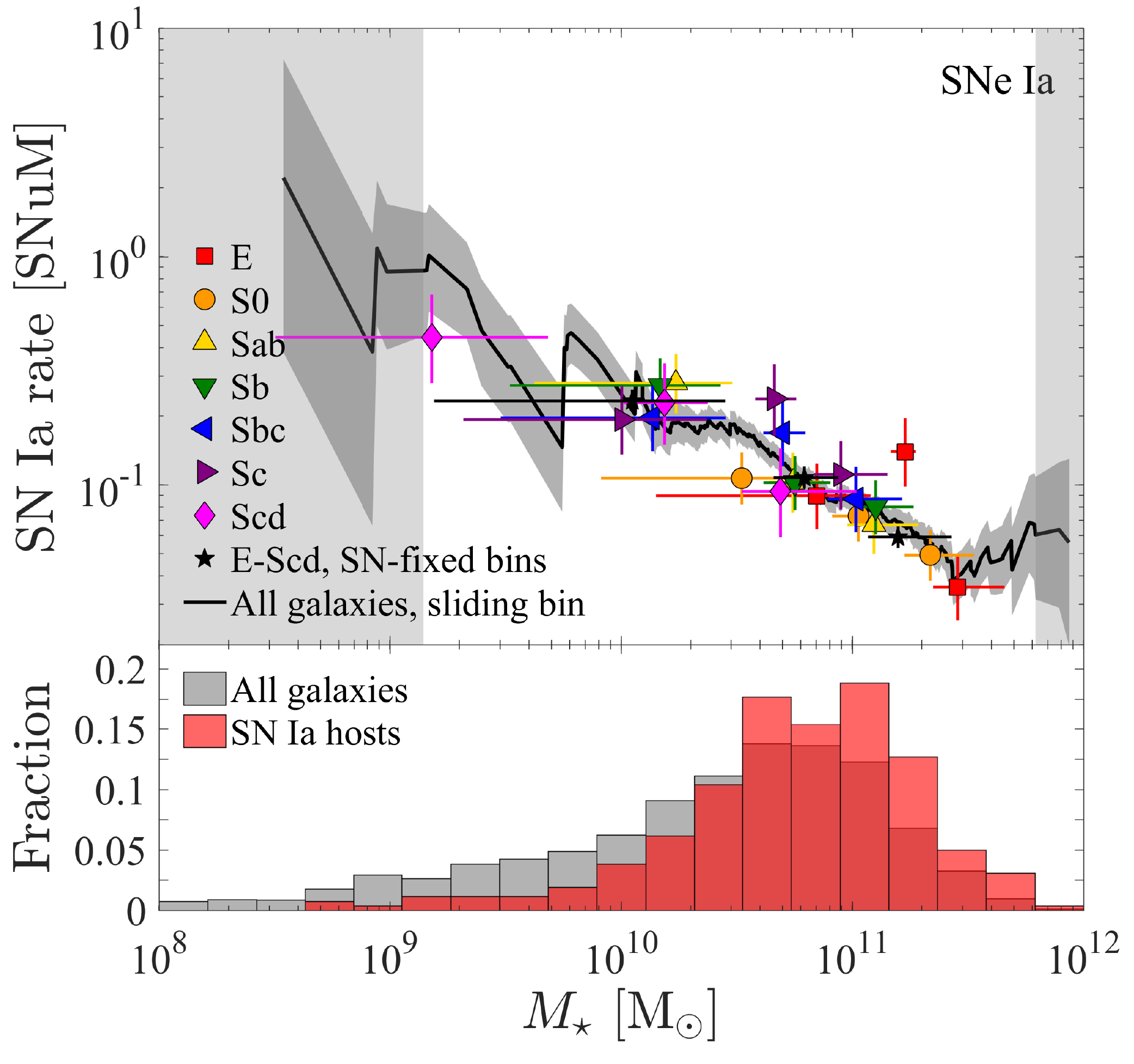} &
  \includegraphics[width=0.47\textwidth]{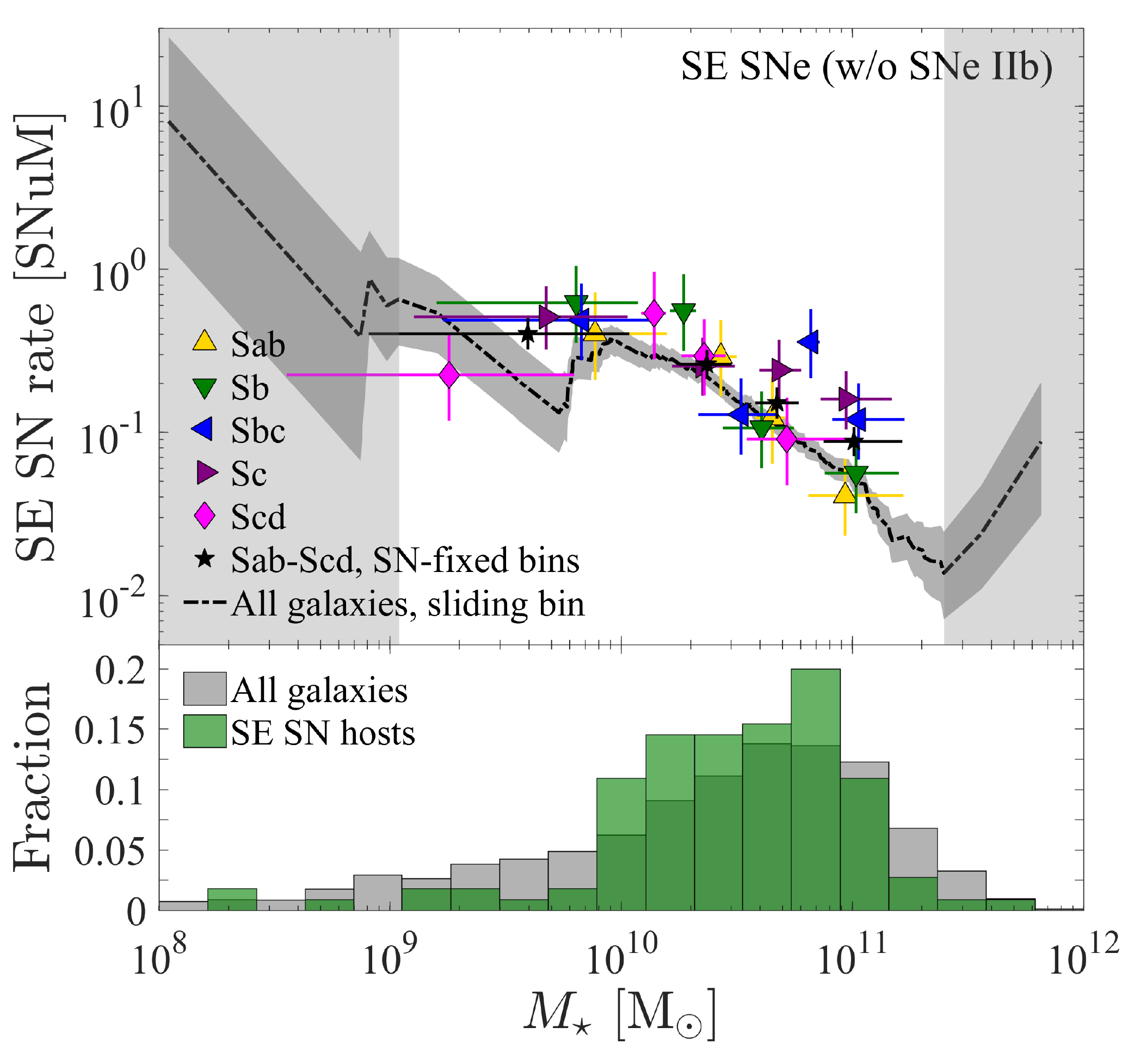} \\
  \multicolumn{2}{c}{\includegraphics[width=0.47\textwidth]{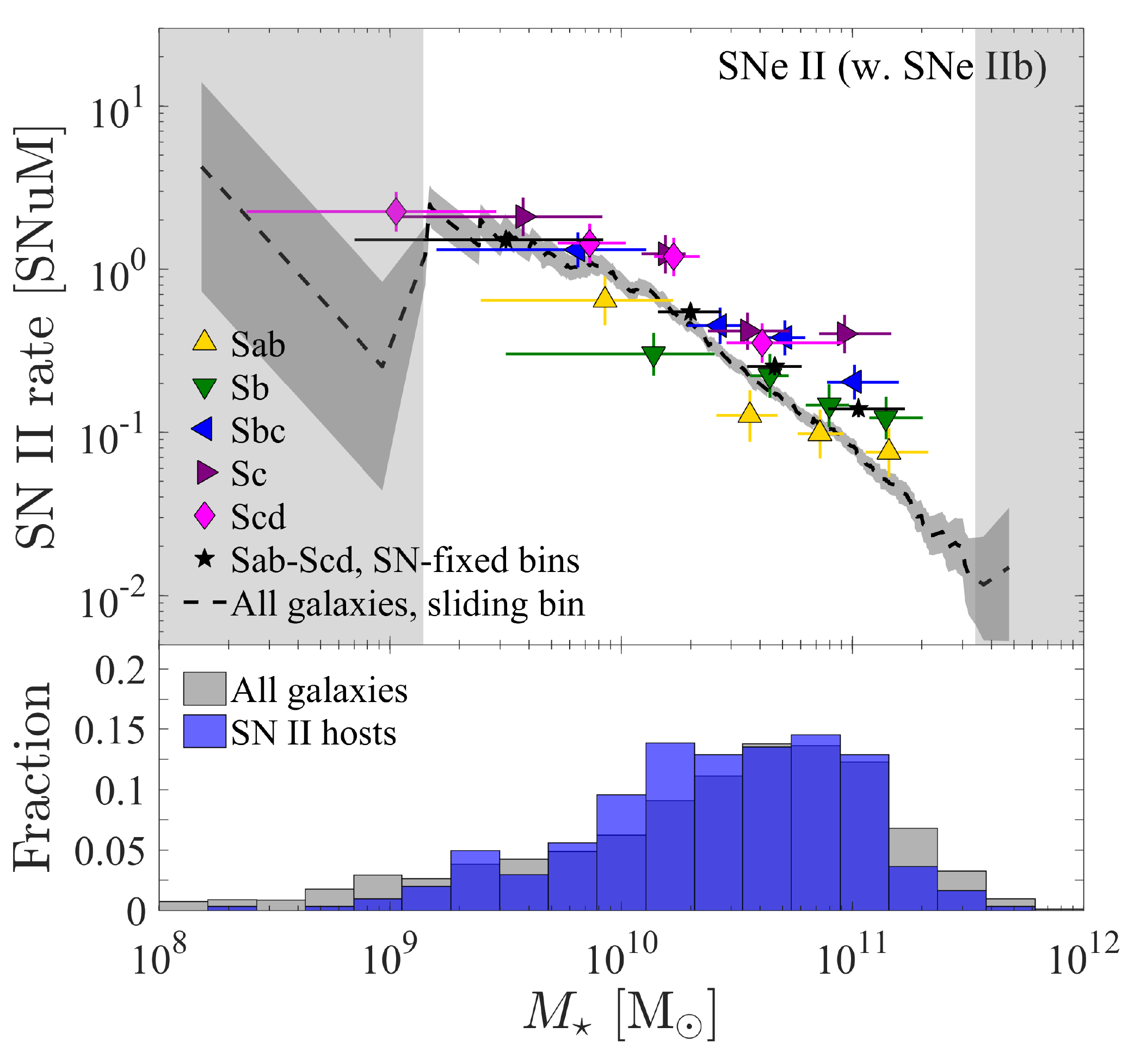} } 
 \end{tabular}
 \caption{Specific SN rates as a function of galaxy stellar mass in galaxies of different Hubble type (symbols, as marked) for SNe Ia (upper left), SE SNe (upper right), and SNe II (bottom center). Only galaxy types that have at least 10 SNe are shown. The curves indicate the rates as measured for all galaxy types using a sliding bin. The 68\% Poisson uncertainties in these rates are shown as the gray shaded regions. Light-gray patches denote regions where the sliding-bin rates are based on $\le 3$ SNe per bin, leading to large Poisson uncertainties. The lower panel of each figure gives the mass distribution of the SN host galaxies (color) superimposed on that of the LOSS galaxy sample (gray). Compare to figures 3 and 4 from L11, where the different rates were scaled to match the rates in Sbc galaxies.}
 \label{fig:rates_all}
\end{figure*}

Fourteen SNe had no spectroscopic classification. \citet{2011MNRAS.412.1419L} divided these SNe into the three SN categories ``according to the statistics of the SNe having spectroscopic classification'' and used them in the L11 rate calculations. We choose to exclude these SNe from the rate measurements performed in this work.

\citet{2011MNRAS.412.1419L} divided the LOSS galaxies into eight Hubble types---E, S0, Sa, Sb, Sbc, Sc, Scd, and Irr (irregular)---according to their designations in the NASA Extragalactic Database.\footnote{\url{http://ned.ipac.caltech.edu/}} We keep these designations in the current work. 

We divide the complete LOSS galaxy sample into several subsamples, which we label the ``LOSS,'' ``sSFR,'' and ``metallicity'' samples. The LOSS sample is similar to the sample used by L11 to measure the original LOSS SN rates (i.e., the ``full-optimal'' subsample; 10,121 galaxies from the targeted list only, i.e., excluding SNe discovered in background galaxies) and is used here to measure SN rates per unit mass (``specific'' rates) as a function of galaxy stellar mass. The sSFR sample comprises 2415 galaxies targeted by LOSS that had spectra acquired by the SDSS and were analyzed by both the MPA-JHU Galspec pipeline\footnote{\url{http://www.sdss3.org/dr9/algorithms/galaxy.php}} \citep{2003MNRAS.341...33K,2004MNRAS.351.1151B,2004ApJ...613..898T} and the NASA-Sloan Atlas\footnote{\url{http://www.nsatlas.org/}} (NSA), which is based on the SDSS DR8 spectroscopic catalog \citep{2011ApJS..193...29A}. This sample is used to correlate the SN rates with sSFR. The ``metallicity'' sample is a subset of 1000 galaxies from the MPA-JHU Galspec catalog, which we use to measure specific SN rates as a function of the central oxygen abundances of the galaxies. We detail the selection criteria for each of the subsamples in Appendix~\ref{appendix:samples}. In Appendix~\ref{appendix:consistency}, we show that the rates measured with each sample are mutually consistent. The numbers of galaxies and SNe in each subsample are summarized in Table~\ref{table:sample_general} and the properties of the galaxies in the entire LOSS sample are given in Table~\ref{table:gal_prop}.


\section{Addenda to the Original LOSS Rates}
\label{sec:addendum}

Before delving into correlations between SN rates and galaxy properties, we use this section to present several updates and extensions to the original LOSS papers. Figures 3 and 4 of L11 show measurements of specific SN rates as a function of galaxy stellar mass in galaxies of different Hubble types. However, in these figures, the rates for each type are scaled to match the rates as measured in Sbc galaxies. This means that the scatter of the rates in different galaxy types is not visible in the figures (though it can be reconstructed to some degree from the parameters of the fits to the original, unscaled measurements, which appear in table 4 of L11), and the rate--mass correlation appears tighter than it really is. In Figure~\ref{fig:rates_all}, we show similar measurements, but do not scale them, so that their spread is apparent. In the SN II rates, where the scatter is largest, there is a clear progression from late-type (i.e., younger and less massive) galaxies, where the rates are highest, to early-type (older, more massive) galaxies. In the SE SN and SN Ia rates, where the scatter is smaller, this trend is not as clear. 

We publish our newly calculated LOSS rates in Table~\ref{table:rates}. L11 fit power laws to their rates and published the fits in their Table 4. They did not, however, publish the measurements themselves. Our rates are not identical to those measured by L11, because we use the entire full-optimal subset of galaxies, after filling in missing masses (see Appendix~\ref{appendix:samples}), but leave out the 14 SNe without spectroscopic classifications (see Section~\ref{sec:galaxies}, above). However, the differences between the LOSS sample used here and the one used by L11 are minor and have little effect on the resultant rates.

L11 divided their SNe into bins so that each rate bin contained roughly the same number of SNe (``SN-fixed'' bins). This is a common practice in studies where the SN sample sizes are either small to begin with or reduced owing to binning (in this case, by binning the host galaxies according to Hubble type); we will follow this binning scheme in the following sections. Once the sample size is increased, though, other binning schemes become available, such as using bins of constant stellar mass (``mass-fixed'' bins). It then becomes necessary to formulate an objective, data-driven method to determine the binning scheme that will extract the most information from the sample. 

To better trace the correlation between the SN rates and stellar mass, we use a sliding mass bin of constant width. The size of this bin depends on the size of the SN sample and is determined by Knuth's rule \citep{2006physics...5197K}, with the added constraint that the bin width not exceed 1 dex of ${\rm M}_\sun$. In each iteration, the bin either gains or loses the nearest SN to its current borders, such that in some iterations, losing one SN on one end may lead to a gain of several SNe on the other. In each bin $i$, the specific SN rate, $R_{i}$, is measured according to
\begin{equation}\label{eq:rate}
 R_{i} = \frac{N_i}{\sum\limits_{j=1}^n t_{c,j} M_{\star,j}},
\end{equation}
where $N_i$ is the number of SNe in the bin, $M_{\star,j}$ is the stellar mass of the $j$th galaxy in the bin, and $t_{c,j}$ is the control (or visibility) time of the $j$th galaxy---i.e., the time during which a given SN type could have been detected by LOSS during the survey. These control times take into account the detection efficiency of the survey, as well as our broad knowledge of SN characteristics (i.e., shapes of light curve and luminosity functions). The resulting rates are reported in Table~\ref{table:rate_snakes}.

We use the original LOSS control times as computed by L11 (see also \citealt{2011MNRAS.412.1419L} and \citealt{li2011LF}; see Section~\ref{subsec:rates_mass} for a discussion of how the inclusion of SNe IIb in the SN II control times may affect the rates measured here). These control times were not included in Tables 2 and 4 of \citet{2011MNRAS.412.1419L}, which laid out most of the properties of the LOSS galaxy and SN samples. In order for our work to be reproducible, and for others to continue to explore the LOSS sample, we publish the control times in Table~\ref{table:sample_general}.

As can be seen in Figure~\ref{fig:rates_all}, the sliding-bin rates diverge wildly at the edges of the mass range in each panel. As the number of galaxies that goes into the denominator in Equation~\ref{eq:rate} grows smaller (as shown in the stellar mass histograms in the bottom panels), and the number of SNe decreases, the Poisson noise dominates, and the rate eventually diverges. This problem also affects rates measured with SN-fixed or mass-fixed bins (as shown in Figure~\ref{fig:all_samples}), but is usually not as apparent.

The sliding-bin rates are useful for visual recognition of possibly interesting features, such as the break and rate ratio discussed in Section~\ref{subsec:rates_mass}, below. However, since they are not independent measurements, they cannot be used for curve fitting. Thus, throughout this work, we use SN-fixed bins for fitting purposes. The number of bins for each fit is chosen to maximize the number of SNe in each bin while also revealing as much structure in the rates as possible. We repeat each fit with a different number of bins to make sure that the results do not change appreciably. All rates in this paper are presented in units of $10^{-12}~{\rm M}_\sun^{-1}~{\rm yr^{-1}}$, which are abbreviated to ``SNuM'' (i.e., SN rate per unit mass). In all the SN rate figures presented here, vertical error bars are 68\% Poisson uncertainties, while horizontal error bars denote the 16th and 84th percentiles of the distribution of galaxies within each bin.

\begin{figure*}
 \begin{tabular}{cc}
  \includegraphics[width=0.47\textwidth]{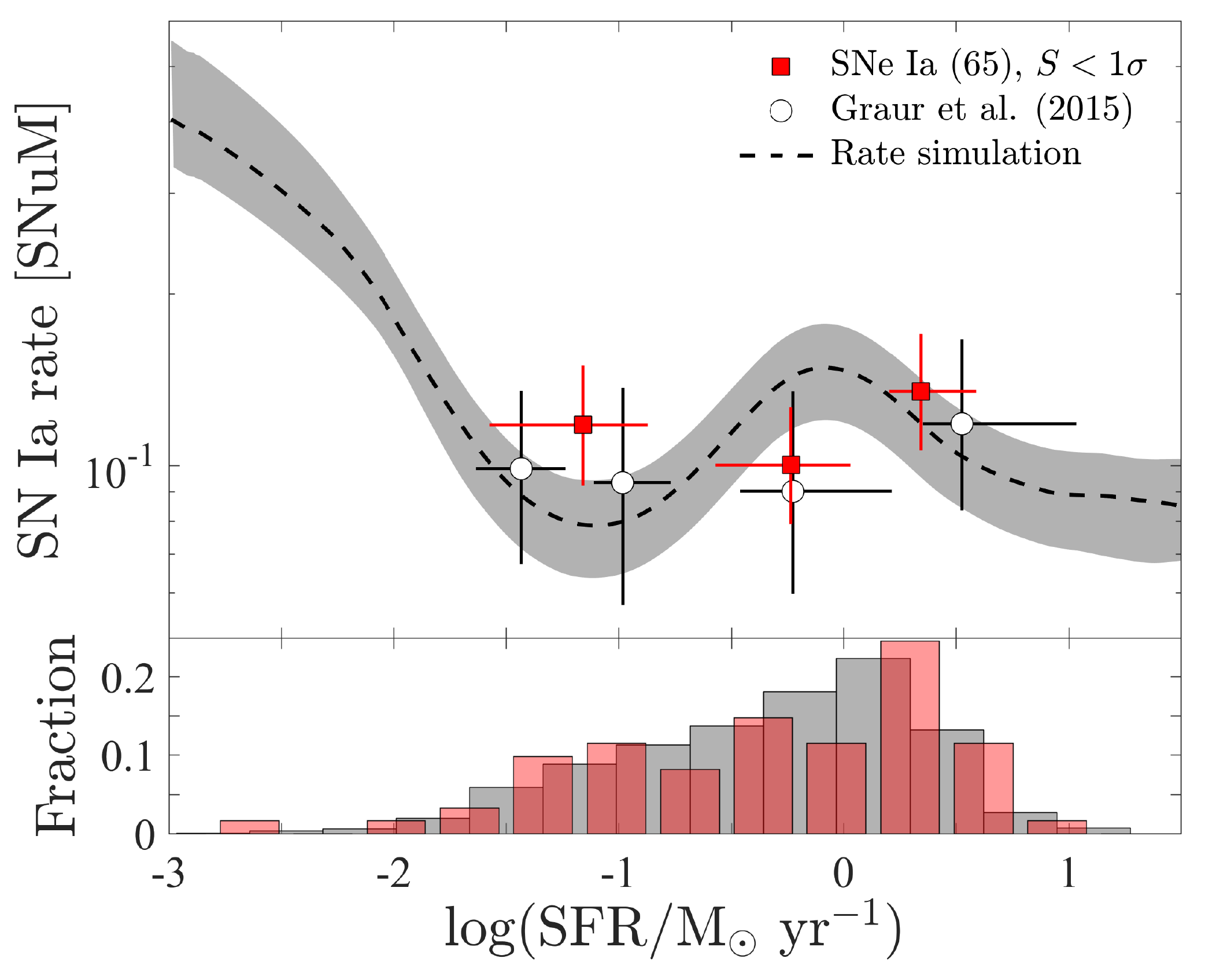} &
  \includegraphics[width=0.47\textwidth]{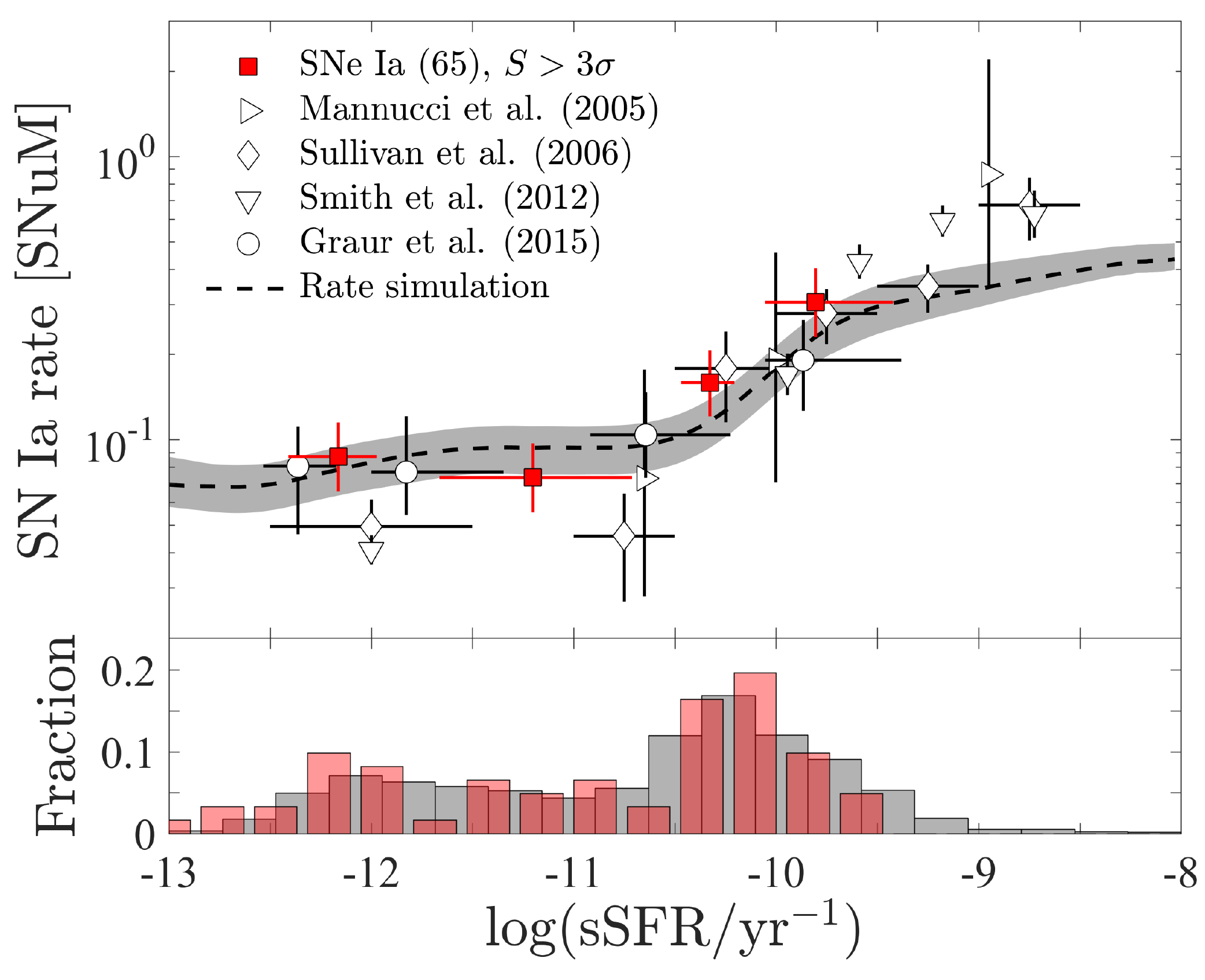}
 \end{tabular}
 \caption{Specific SN Ia rates as a function of SFR (left) and sSFR (right). The rates measured here (red squares) are consistent with previous measurements from the literature, as marked by white symbols (\citealt{2005A&A...433..807M,2006ApJ...648..868S,2012ApJ...755...61S}; G15), as well as with the G15 model that combines the SN Ia DTD with the scaling relation between galaxy age and stellar mass (dashed curve with gray 68\% uncertainty region, reproduced from G15). As in Figure~\ref{fig:rates_all}, the bottom panels show distributions of the galaxy property in question in the SN host galaxies (color) and the entire sample (gray).}
 \label{fig:sSFR_Ia}
\end{figure*}

\begin{figure}
 \includegraphics[width=0.47\textwidth]{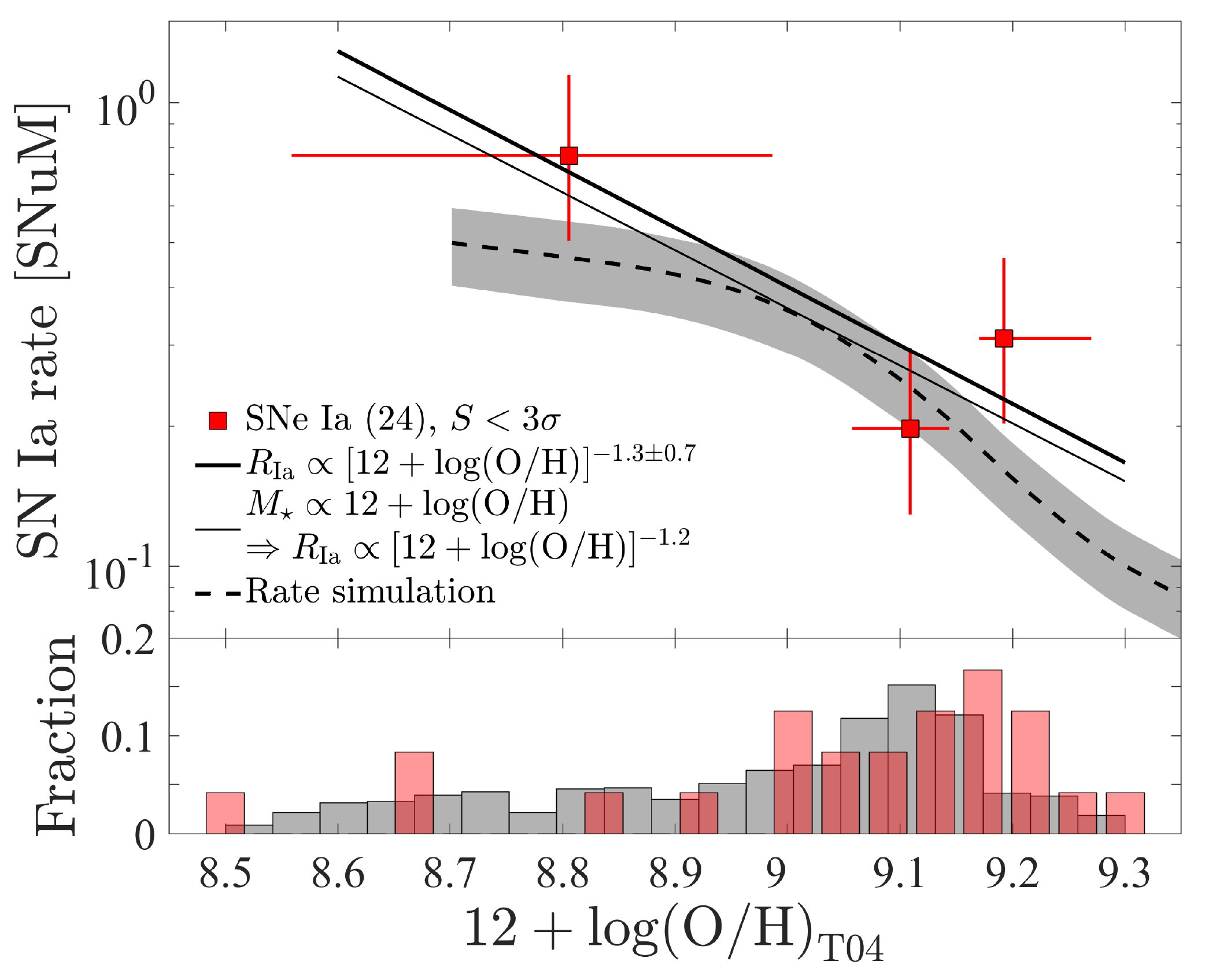}
 \caption{Specific SN Ia rates as a function of galaxy metallicity. A first-order polynomial fit to the data is shown as a thick curve (though formally, there is no significant correlation between the rates and metallicity). The rate--mass correlation, coupled with the galaxy scaling relation between $M_\star$ and metallicity, is consistent with the measurements (thin curve), as is the G15 rate simulation, rebinned according to the metallicity values of the galaxies in the metallicity sample (dashed curve with gray 68\% uncertainty region). SDSS metallicity values were measured using the T04 metallicity scale.}
 \label{fig:OH_Ia}
\end{figure}

\begin{figure}
 \includegraphics[width=0.47\textwidth]{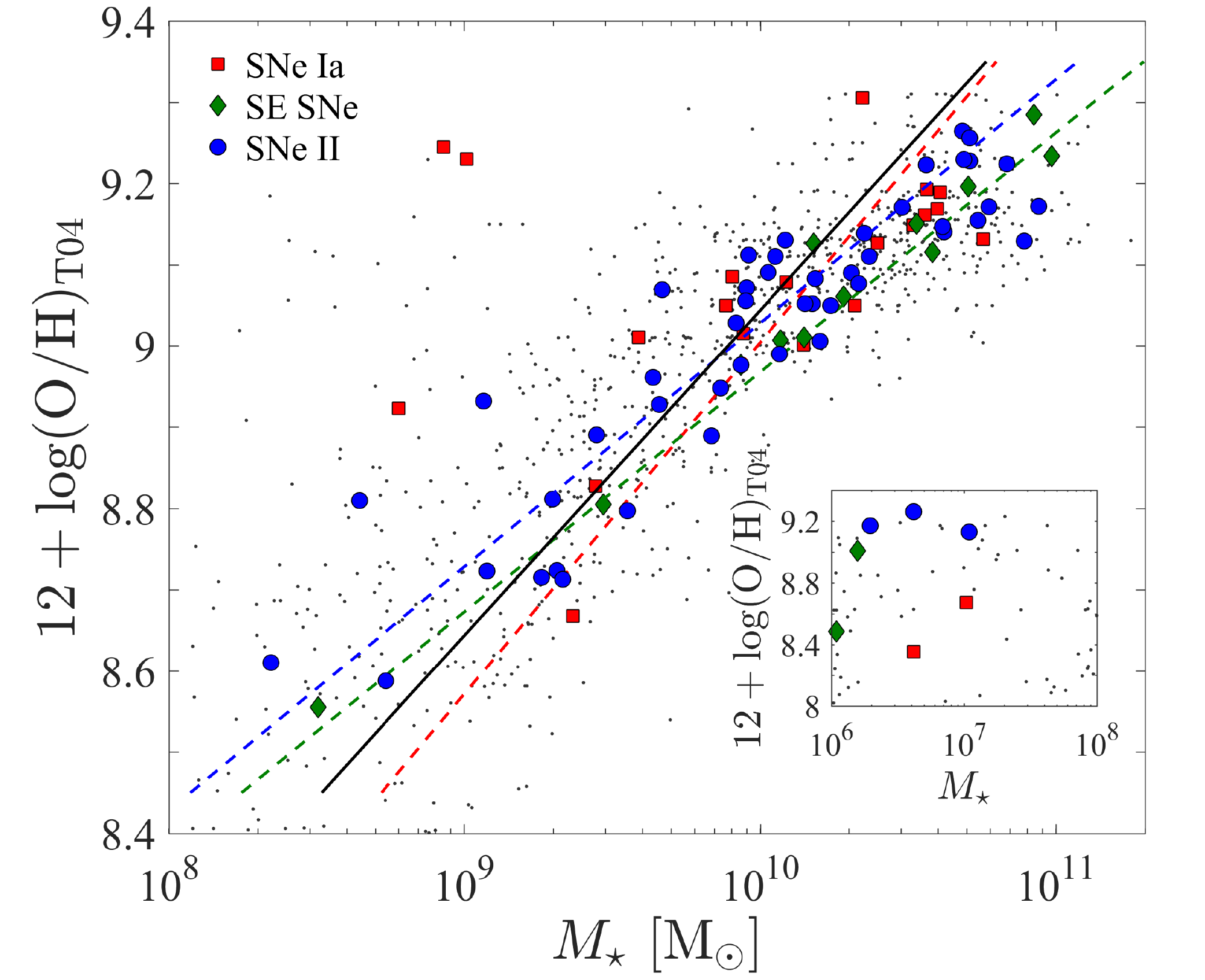}
 \caption{Galaxy metallicity vs. stellar mass. Gray squares represent the 1000 SDSS galaxies in the metallicity sample, while colored symbols represent SN host galaxies: SNe Ia (red squares), SE SNe (green diamonds), and SNe II (blue circles). The curves are linear fits to these datasets in log--log space. The solid line is a fit to all 1000 galaxies, while the dashed lines are fits to the separate SN host samples. The line colors match those of the symbols. SDSS metallicity values were measured using the T04 metallicity scale.}
 \label{fig:OH}
\end{figure}


\section{SN Ia Rates}
\label{sec:SNIa}

We fit the SN Ia rates as a function of galaxy stellar mass shown in Figure~\ref{fig:rates_all}, in all galaxies combined, with a linear fit (i.e., a first-order polynomial) in log--log space (not shown). The resultant slope, $-0.44 \pm 0.06$, is consistent with that found by L11, $-0.50 \pm 0.10$. We also find that a suspected bend in the SN Ia rates at $\sim5\times10^{10}~{\rm M_\sun}$, seen in Figure~\ref{fig:rates_all}, is not statistically significant, and conclude that the SN Ia rate evolves smoothly over the range of galaxy mass probed here. 

Several studies have now shown an anticorrelation between specific SN Ia rates and host-galaxy stellar masses (\citealt{2006ApJ...648..868S}; L11; \citealt{GraurMaoz2013}). However, all of these studies noted that while the anticorrelation was significant for star-forming galaxies, it was unclear whether it persisted in passive galaxies as well. Using the likelihood-ratio test (see Appendix~\ref{appendix:ratio}), we find that while there is a significant trend in spiral galaxies ($S>5\sigma$), there is only a $>3\sigma$ trend in E+S0 galaxies (as well as in S0 galaxies alone) and a trend with $2\sigma<S<3\sigma$ (i.e., insignificant) trend in E galaxies. These findings are in agreement with L11. Furthermore, we measure SN Ia rates in star-forming (${\rm log(sSFR/yr^{-1})}>-12$) and passive (${\rm log(sSFR/yr^{-1})}\leq-12$) galaxies using the sSFR sample. We find a $>5\sigma$ trend between the SN Ia rates and galaxy stellar mass in the former, but no trend ($S<2\sigma$) in the latter. Our sample includes only 35 SNe Ia in E galaxies and 22 in passive ones, so the absence of a significant trend may be due to small-number statistics. Although the LOSS galaxy sample is biased toward high-mass galaxies, which restricts most of the E and passive galaxies to a narrow mass range of about $10^{10}$--$5\times10^{11}~M_\sun$, this should not be the reason why we find no significant trend, because S0 galaxies (with 56 SNe Ia)---for which there is a significant trend---occupy the same mass range.   

In Figure~\ref{fig:sSFR_Ia}, we show specific SN Ia rates as a function of SFR and sSFR. The rates as a function of SFR shown here are consistent with those from G15. As in that work, the new rates favor a flat trend, but owing to their large statistical uncertainties they are also consistent with the G15 model rates. The G15 model could be challenged either by measuring more precise rates or by targeting low-mass, passive galaxies, where the G15 model predicts rates $\sim 3$--4 times higher than the flat trend.

Formally, the second set of measurements can be fit with a simple linear function, at a significance of $>3\sigma$. However, it is more interesting to note that these measurements follow the same pattern observed by previous studies (\citealt{2005A&A...433..807M,2006ApJ...648..868S,2012ApJ...755...61S}; G15): the rates are flat in passive galaxies and rise with rising sSFR values in star-forming galaxies. In G15, we showed that this pattern is another result of the interplay between the SN Ia DTD and galaxy scaling relations. 

In Figure~\ref{fig:OH_Ia}, we show specific SN Ia rates vs. metallicity. There is no significant correlation, but, if we replace $M_\star$ in the SN Ia rate--mass correlation with metallicity via the galaxy scaling relations shown in Figure~\ref{fig:OH}, the resultant correlation is consistent with the measured rates. Rebinning the G15 rate simulation\footnote{A combination of a $t^{-1}$ DTD and the correlation between galaxy age and stellar mass; see \citet{GraurMaoz2013} for more details.} according to the metallicity values of the galaxies shows that our favored model is also broadly consistent with the measurements.


\section{Core-collapse Supernova Rates}
\label{sec:CCSN}

In this section, we show that the SE SN rates are depressed, relative to the SN~II rates, in low-mass galaxies. This result, which was hinted at in L11, is more significant than first thought. We attempt to explain this result by investigating possible correlations between the SN rates and other galaxy properties (sSFR and metallicity) and show that any correlation between the SN rates and one of the galaxy properties examined here can be transformed, with the aid of previously known galaxy scaling relations, into the measured correlation with the other two galaxy properties.

\subsection{CC SN Rates and Galaxy Stellar Mass}
\label{subsec:rates_mass}

In Figure~\ref{fig:rates_all}, we show specific CC SN rates as a function of galaxy stellar mass. Using the likelihood-ratio test, we find a break (i.e., a ``knee'' in the curve) at $\sim 10^{10}~{\rm M}_\sun$ in both the SN II and SE SN rates, as measured with all galaxy types. This break has a significance of $>3$--5$\sigma$, depending on how many bins are used to calculate the rates. However, this break is no longer statistically significant when the galaxy sample is limited to Sab--Scd galaxies, so we attribute it to the inclusion of E and S0 galaxies in the rate calculation. These galaxies, which tend to be massive, passive galaxies, do not host CC SNe and thus drag down the rates in the high-mass range. The lack of a break in the CC SN rates in star-forming galaxies implies a smooth dependence of the rate on either SFR or a combination of galaxy properties, such as SFR and metallicity.

The sliding bin used to calculate the rates shown in Figure~\ref{fig:rates_all} also makes it appear as if there is a dip in the SE~SN rates at $\sim 6\times10^9~{\rm M}_\sun$. We attribute this to Poisson noise stemming from the width of the bin. This dip disappears in Figure~\ref{fig:all_samples}, where we use different techniques to measure the rates in discrete bins.

We measure the slopes of the SE SN and SN II rates as a function of galaxy stellar mass and find that the SE SN rates have a shallower decline with stellar mass than the SN II rates: $-0.46 \pm 0.10$ ($-0.64 \pm 0.09$) as opposed to $-0.68 \pm 0.05$ ($-0.84 \pm 0.05$), as fit when using Sab--Scd (all) galaxies. L11, on the other hand, found that the SE SN and SN II rates had identical slopes ($0.55\pm0.10$).

\begin{figure}
 \center
 \includegraphics[width=0.47\textwidth]{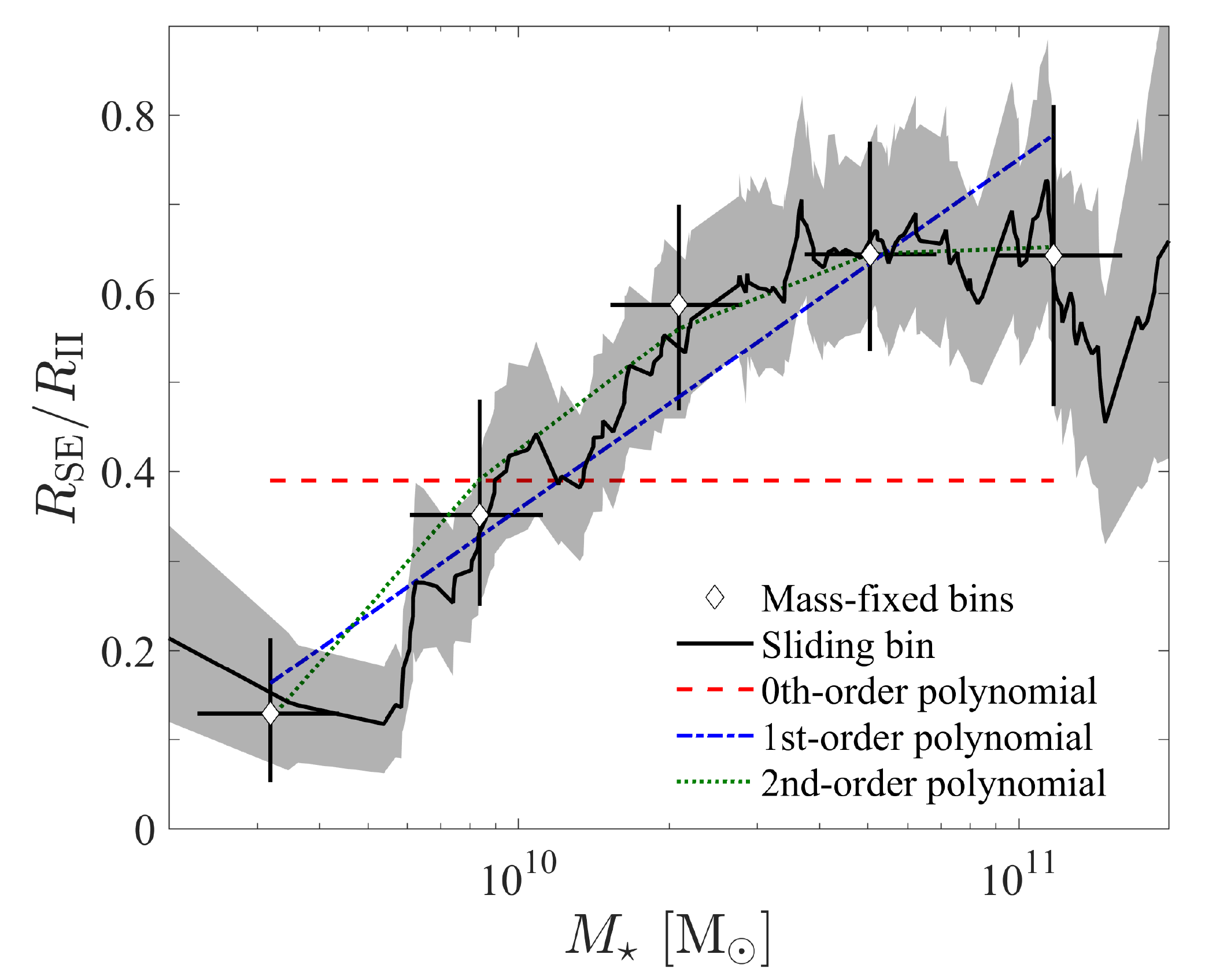}
 \caption{The ratio between the SE SN and SN II rates. The diamonds mark the rate-ratio measurements in bins with roughly equal ${\rm log}(M_\star)$ values. The black curve indicates the rate-ratio measured by interpolating the SN II rates measured with a sliding bin to those of the SE SN rates and is shown here only to guide the eye. The shaded region, as well as the vertical error bars, represent the 16th and 84th percentiles of the distribution that results from dividing the Poisson distributions of the numbers of SE SNe and SNe II in each bin. The red dashed, blue dot-dashed, and green dotted curves are zeroth-, first-, and second-order polynomial fits to the mass-binned measurements. The likelihood-ratio test for these fits shows that the first-order polynomial is a better fit than the zeroth-order polynomial at a $>4\sigma$ significance; the second-order polynomial is a better fit than the zeroth-order polynomial at a $>3\sigma$ significance. However, the likelihood ratio test does not favor either the 1st- or 2nd-order polynomial (they are consistent with each other at a $<2\sigma$ significance level).}
 \label{fig:likelihood_ratio}
\end{figure}

Although the values of the slopes depend on which galaxy sample we use to derive the rates, the \emph{ratio} of the SE SN to SN II rates, $R_{\rm SE}/R_{\rm II}$, shown in Figure~\ref{fig:likelihood_ratio} and collected in Table~\ref{table:rate_ratio}, has the same structure: it rises from 0.2 to 0.6 in galaxies with masses $\la 2\times 10^{10}~{\rm M}_\sun$, then remains constant. L11 made a similar measurement of the ratio of the SE SN rates to the total CC SN rate. They noted that the lowest-mass measurement was lower than the others, but that this was only a $2\sigma$ effect. As shown in Figure~\ref{fig:likelihood_ratio}, a likelihood-ratio test shows that either a first- or second-order polynomial provides a better fit to $R_{\rm SE}/R_{\rm II}$ than a constant (i.e., zeroth-order polynomial, which represents the possibility that there is no trend), at a significance of $>3\sigma$. Varying the number of bins and the mass range over which the test is performed results in similar significance values. Altogether, this means that the trend we see in the ratio between the SN rates is not simply a result of the uncertainties of the measurements, but a real effect. We note that L11 included systematic uncertainties in their rates, while we only consider statistical uncertainties. However, as the SE~SN and SN~II rates should suffer from similar systematics, the ratio between the rates should only be affected by the statistical uncertainties.

\begin{figure}
 \includegraphics[width=0.47\textwidth]{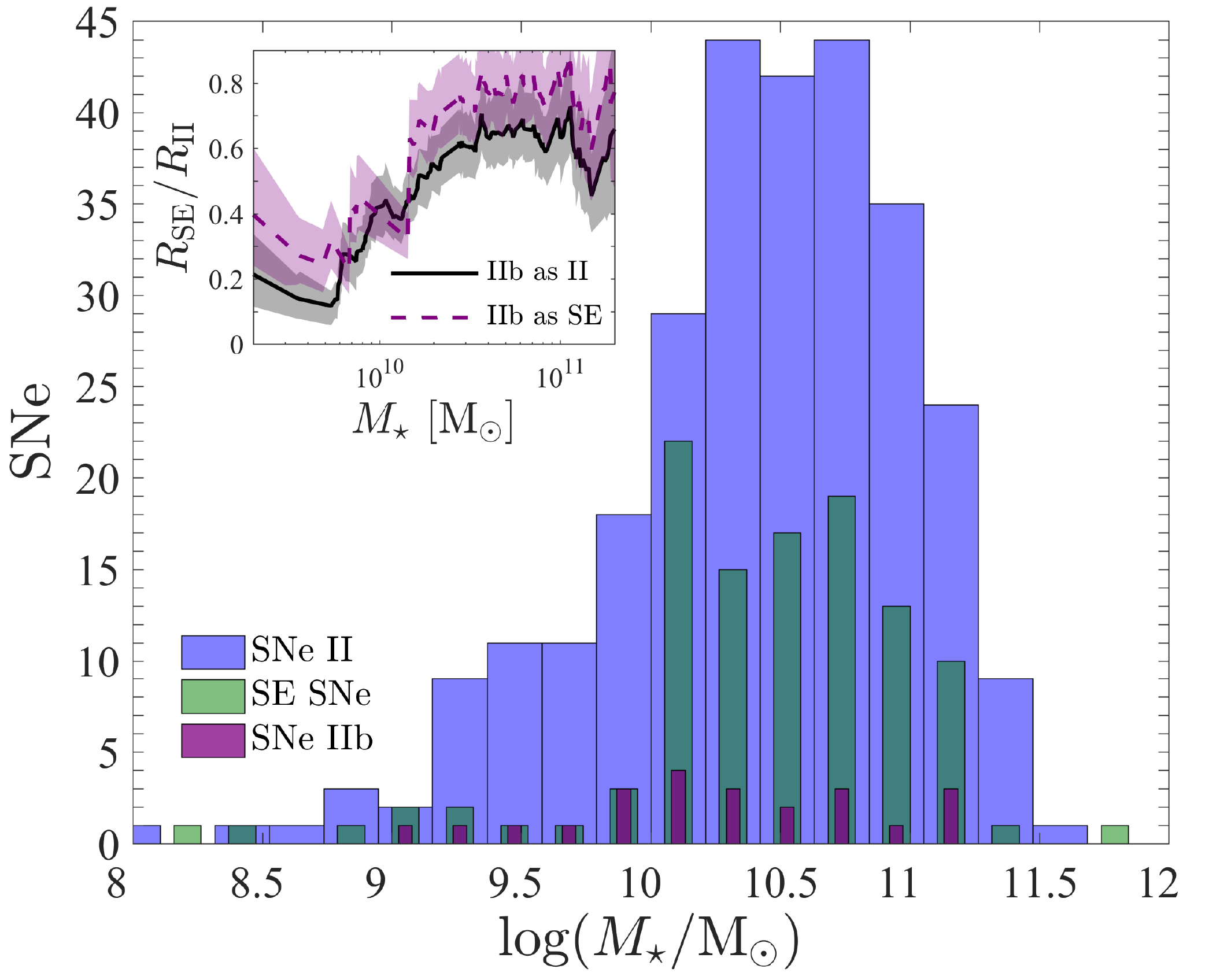}
 \caption{The distribution of SN IIb host-galaxy stellar masses (purple) plotted against those of SE~SN (green) and SN II host galaxies (without SNe~IIb; blue). While adding the SNe~IIb in galaxies with stellar masses $\leq10^{10}~{\rm M}_\sun$ would roughly double the SE SN sample in the range $10^{9}<M_\star<10^{10}~{\rm M}_\sun$, there are still roughly five times more SNe II in that range. The inset shows the resultant ratios between SE~SN and SN~II rates when the SNe IIb are treated as SNe II (black solid) or as SE~SNe (purple dashed). While the scaling of the latter ratio is slightly higher, the general trend---lower SE~SN rates relative to the SN~II rates in galaxies with $M_\star<10^{10}~{\rm M}_\sun$---is still present.}
 \label{fig:IIb}
\end{figure}

When they computed the SN control times, L11 treated SNe IIb as SNe II, though they are considered to be a type of SE SN (e.g., \citealt{1988AJ.....96.1941F,1997ARA&A..35..309F,1993ApJ...415L.103F}). In this work, we have chosen not to recalculate the control times, so our SN II rates remain contaminated by SNe~IIb. To test whether this is the reason for the deficiency of SE SNe relative to SNe II in low-mass galaxies, we remeasure the SN rates, using the same control times but with SNe IIb removed from the SN II sample and added to the SE SN sample. The resulting rates will not be strictly correct, but will reveal whether the addition of the SNe IIb to the SE SN sample can make up for the deficit. In Figure~\ref{fig:IIb}, we show that the number of SNe IIb in low-mass galaxies, while of the same order of magnitude as other SE SNe, is not enough to make up for the deficiency of SE SNe in low-mass galaxies relative to SNe II.

\begin{figure}
 \includegraphics[width=0.47\textwidth]{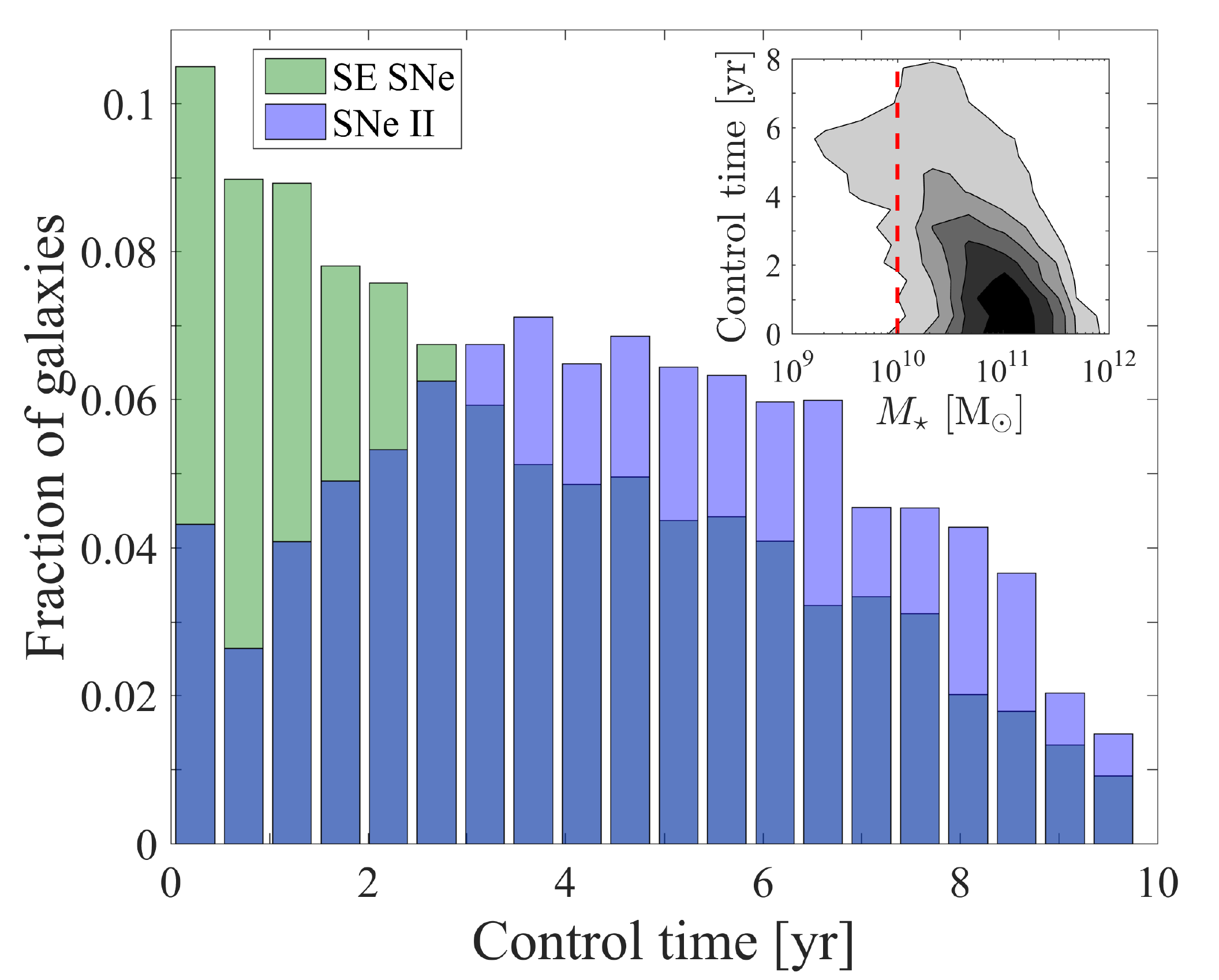}
 \caption{The distribution of SE SN (green) and SN II (blue) control times in the LOSS sample. The inset shows the distribution of SE SN control times as a function of galaxy stellar mass. The red dashed line marks a threshold of $10^{10}~{\rm M}_\sun$. In galaxies below this threshold, the control times are longer (i.e., the survey is more sensitive to SE SNe) and are clustered in the range 4--6 yr. In this range the SN II and SE~SN control times are very similar, with a slight advantage for SNe II, which have control times larger by roughly a third. This difference is not enough to offset the order of magnitude difference in the number of SE SNe and SNe II discovered in low-mass galaxies. }
 \label{fig:ct}
\end{figure}

The deficiency of SE SNe in low-mass galaxies could also be explained by a bias on the part of the LOSS survey toward this SN type, but we have no reason to think that such a bias exists. First, all SNe should be easier to discover in low-mass (and hence lower-luminosity) galaxies, where the contrast between the SN and galaxy light is greater than in more massive galaxies. Second, on average, SE SNe are more luminous than SNe II (e.g., L11b; \citealt{2011ApJ...741...97D,2012ApJ...744...10K,2014AJ....147..118R}). At the same time, they brighten and decline faster than SNe II. The latter are dominated by SNe IIP, which have a long plateau phase of $\sim 100$ days shortly after explosion. These properties are accounted for in the control times, which end up being very similar for SE SNe and SNe II. We show this similarity in Figure~\ref{fig:ct}, where the control times in galaxies with $M_\star \la 10^{10}~{\rm M}_\sun$, which lie in the range 4--6 yr, are very similar for both SN types. The SN II control times are larger by $\sim 1/3$, which means the survey is more sensitive to them in these low-mass galaxies. However, this small difference is not enough to balance the factor $\sim 10$ difference between the number of SNe II and SE SNe in these galaxies, as seen in Figure~\ref{fig:IIb}. 

Finally, we must take into consideration the targeted nature of LOSS. This survey targeted large, massive galaxies, so that while it is complete down to galaxies with absolute $K$-band magnitudes of $-24$, it is deficient in low-luminosity galaxies. So, if SE SNe preferentially explode in low-luminosity galaxies, the LOSS sample might be biased against them. However, in this work we measure the ratio between the SE SN and SN II rates, which gradually declines in galaxies with $M_\star \lesssim 10^{10}~{\rm M}_\sun$. Thus, even if the incompleteness of the galaxy sample were to translate into a systematic effect on the SN rates, that effect should be identical for all SN types and should cancel out (for why would LOSS know to preferentially select low-luminosity galaxies that hosted one kind of SN over another?).

\subsection{CC SN Rates and Star Formation Rates}
\label{subsec:SFR}

In Figure~\ref{fig:rates_CC_sSFR}, we plot specific SN rates as a function of galaxy sSFR (reported in Table~\ref{table:rates_OH}). Our measurements are consistent with the SN II rates from G15 and combined CC SN rates from \citet{botticella2012}. While we find no CC SNe in passive galaxies, we note that there are one SE SN and 11 SNe II in galaxies with very low sSFRs: ${\rm log(sSFR/yr^{-1})}<-11$. As we explain in Appendix~\ref{appendix:samples}, this may be due to a known problem with the SDSS pipeline, which ``shreds'' large angular-size galaxies into multiple components. While we have corrected for this by measuring our own sSFRs from NSA photometry, the largest (and hence most massive and with the lowest sSFR) galaxies may still suffer from some residual shredding. Thus, the polynomial fits shown in Figure~\ref{fig:rates_CC_sSFR} were only fit to the measurements in galaxies with ${\rm log(sSFR/yr^{-1})}\ge-11$. The resultant fits are consistent with the low-sSFR measurements as well. We find slopes of $0.8 \pm 0.5$ for SE~SNe and $0.9 \pm 0.2$ for SNe II. These values are consistent with the slope of $1.33^{+0.41}_{-0.35}$ found by G15 for SNe II. The SN II slope is also consistent with a slope of 1. Since the progenitors of SNe IIP, which make up the bulk of SNe II, are red supergiants, we expect the SN II rate to be directly proportional to the sSFR of the stellar population to which it belongs. 

The SN II and SE SN rate slopes are consistent with each other. Formally, though, the SE SN rates have a slightly shallower slope than the SN II rates. If this difference between the slopes was confirmed with a larger SN sample, it would be in line with the trend we observe in low-mass galaxies, which are, on average, more star-forming. This should be confirmed with a larger SN sample. We also note that in G15 we showed that the rate--mass correlation could be transformed into the rate-sSFR correlation by plugging in the galaxy scaling relations between stellar mass and sSFR. This remains true here.

\subsection{CC SN Rates and Metallicities}
\label{subsec:OH}

In Figure~\ref{fig:rates_OH}, we show specific SN rates as a function of galaxy metallicity (collected in Table~\ref{table:rates_OH}). Because these metallicity values are derived from SDSS spectra of the galaxies, they represent the metallicity value in the centers of the galaxies alone. Many studies have shown that galaxies have metallicity gradients, where the metallicity decreases as a function of the distance from the galaxy center (e.g., \citealt{1989epg..conf..377D,1994ApJ...420...87Z,1997ApJ...489...63G,1999PASP..111..919H,2000A&A...363..537R,2012A&A...546A...2S}). Some studies have shown that in certain galaxies, this gradient flattens or even falls off in the centers of the galaxies (e.g., \citealt{1989epg..conf..377D,1992MNRAS.259..121V,2009ApJ...695..580B,2011MNRAS.415.2439R}). This means that the metallicities at the locations of the SNe are bound to be different from those measured from the SDSS spectra and used here \citep{Modjaz2011}. 

\begin{figure}
 \includegraphics[width=0.47\textwidth]{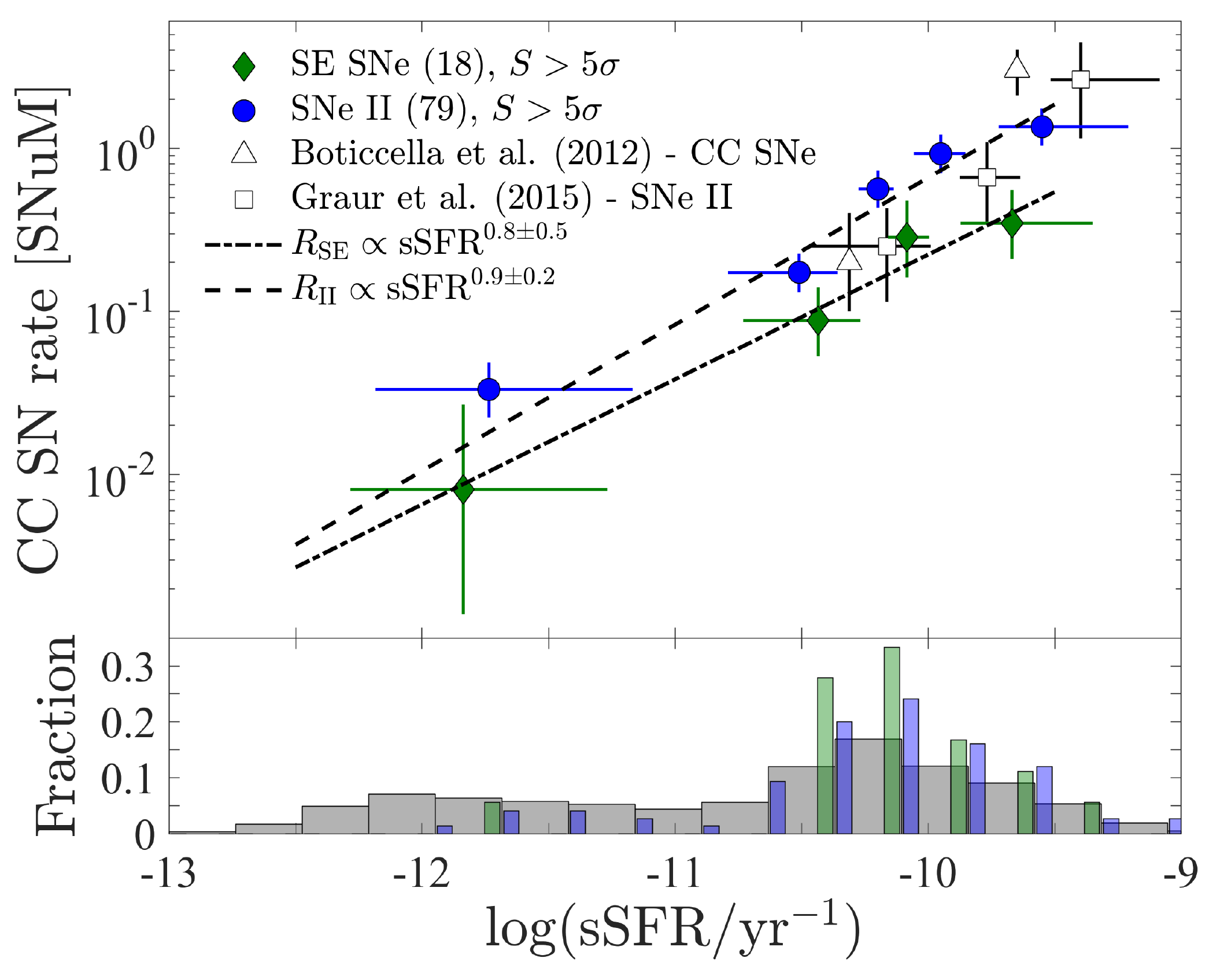}
 \caption{Specific CC SN rates as a function of sSFR. Polynomial fits to the SN II (blue circles) and SE SN (green diamonds) rates are consistent with each other and with a slope of 1, which is expected if all stars above a certain mass threshold end up exploding as members of these SN families. Formally, the SE SN rates have a shallower correlation with sSFR than the SN II rates, echoing the deficiency of SE SNe in low-mass galaxies shown in Figure~\ref{fig:likelihood_ratio}. A larger sample of SNe is required to confirm this trend. For display purposes only, the SN II rate in low-sSFR galaxies has been shifted to the right by 0.1 dex. White triangles denote CC SN rates measured by \citet{botticella2012}. Measurements from G15, which only included SNe~IIP and IIL, are shown as white squares. }
 \label{fig:rates_CC_sSFR}
\end{figure}

\begin{figure}
 \includegraphics[width=0.47\textwidth]{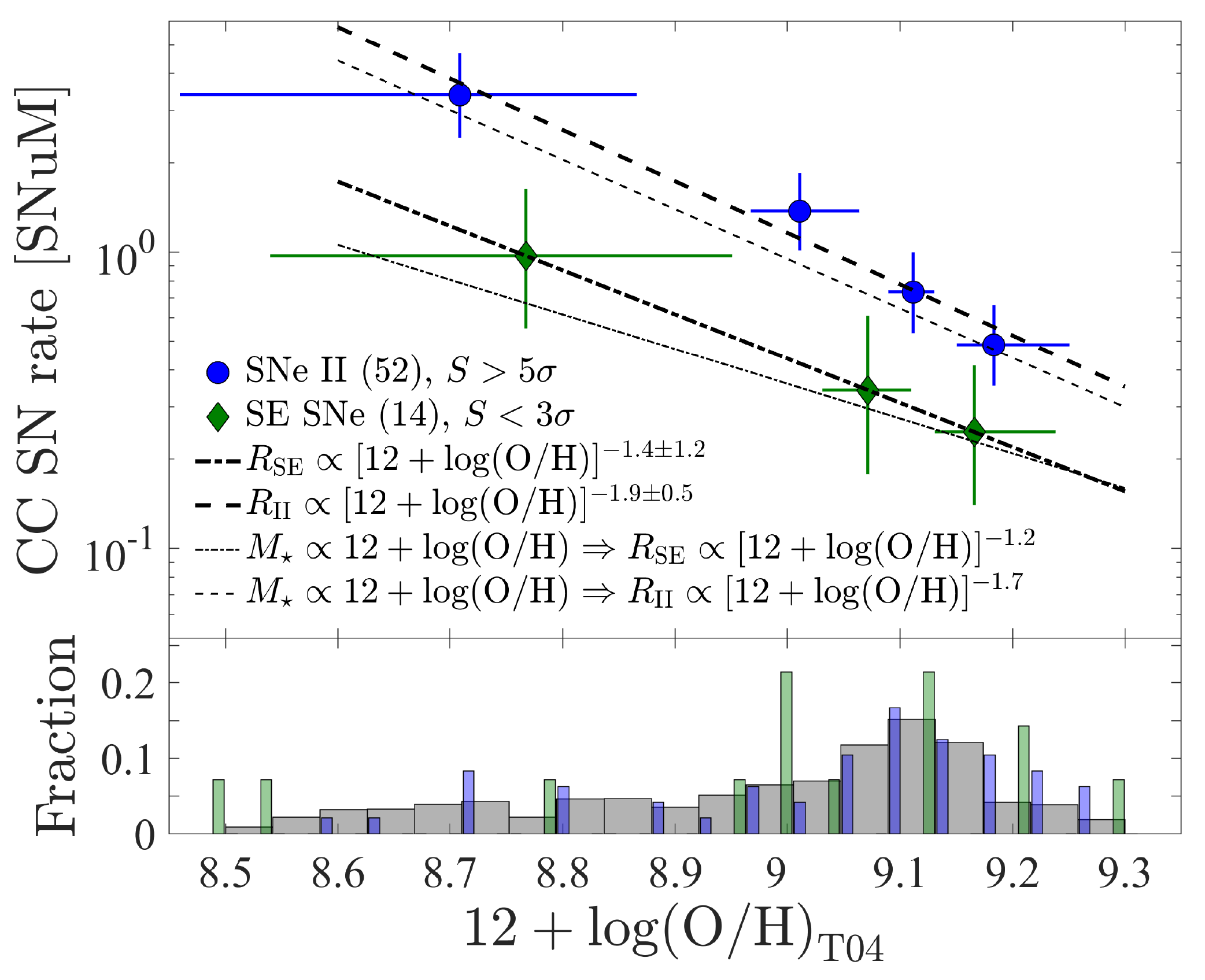}
 \caption{Specific SN rates as a function of galaxy metallicity, as measured in bins with roughly equal numbers of SNe (symbols). The bottom panel shows the distribution of galaxy metallicity, overlaid with similar distribution for host galaxies of SE SNe (green) and SNe II (blue). There is a $>5\sigma$ anticorrelation between the SN II rates and metallicity values, but no statistically significant correlation for SE SNe (thick curves). The correlation of the rates with metallicity can be explained through well-known correlations between galaxy properties. By substituting the correlation between galaxy stellar mass and metallicity shown in Figure~\ref{fig:OH} into the anticorrelation between SN rates and stellar mass, we get the thin black curves, which are consistent with the measurements. SDSS metallicity values were measured using the T04 metallicity scale.}
 \label{fig:rates_OH}
\end{figure}

However, the central metallicity values can still be used as a proxy for the local values, in certain circumstances. Recently, \citet{2016A&A...591A..48G} used integral-field unit spectroscopy of 115 CALIFA (Calar Alto Legacy Integral Field Area Survey; \citealt{2012A&A...538A...8S}) galaxies that hosted 132 SNe to study biases between metallicities measured in different locations and with different fiber apertures. This sample is similar to LOSS in both stellar mass and redshift range. From their Table 5, one can see that central metallicity values are systematically higher than local values. However, these biases are small, $<0.1$ dex (though this specific result may be due to the small dynamical range afforded by the \citealt{2013A&A...559A.114M} metallicity scaling used in that work). This result is consistent with the previous claim by \citet{2012ApJ...758..132S} that central metallicities are equal to local values to within $0.1$ dex (though \citealt{2012ApJ...758..132S} used higher-redshift galaxies with smaller angular sizes than either the LOSS or CALIFA galaxies). 

\citet{2016A&A...591A..48G} found that the best proxy for the local metallicity was a combination of the central metallicity and the metallicity gradient of the galaxy. Measuring the metallicity gradients of the LOSS galaxies is beyond the scope of this paper. Thus, in this work, we use the central metallicities measured by Galsepc as a proxy for the metallicities at the locations of the SNe, but note that the metallicities at the locations of the SNe are expected to be lower than the values used here. 

We find no statistically significant differences between the slopes of the mass-metallicity correlations for the different SN types. On the other hand, we find a significant ($>5\sigma$) anticorrelation between the specific SN II rates and metallicity. Formally, the SE SN rates exhibit a similar anticorrelation with metallicity, but owing to the large uncertainties of the measurements, this trend is not statistically significant. It would be surprising to find any sort of trend between SN~II rates and metallicity, were this correlation assumed to imply a relation. This trend can, however, be easily explained away by combining the galaxy scaling relation between stellar mass and metallicity, shown in Figure~\ref{fig:OH}, and the observation that the SN II rates decrease with increasing galaxy stellar mass, shown in Figure~\ref{fig:rates_all}. When this scaling relation is inserted into the SN~II rate--mass correlation, the result provides a good fit to the rates vs. metallicity. Using scaling relations obtained by fitting only SN host galaxies provides the same result. Thus, the trend between SN II rates and metallicity can be attributed to galaxy scaling relations and not to any intrinsic dependence of SN II rates on metallicity. We discuss this further in Section~\ref{subsec:CCSN_context}, below.

The above holds true for SE SNe as well. While the thick curve in Figure~\ref{fig:rates_OH} shows a formal fit to the data, the thin curve is the result of inserting the galaxy scaling relation into the rate--mass correlation. As in Figure~\ref{fig:rates_CC_sSFR}, while the slopes of the fits to the SE SN and SN II rates hint at a deficiency of SE SNe in low-metallicity (hence, low-mass) galaxies, due to the size of the metallicity sample, this effect is not statistically significant.

The MPA-JHU Galspec metallicity values were measured using the T04 scale, which is based on the [O II] $\lambda3727$, H$\beta$, [O III] $\lambda5007$, [N II] $\lambda\lambda6548,~6584$, and [S II] $\lambda\lambda6716,~6731$ emission lines \citep{2004MNRAS.351.1151B}. \citet{2008ApJ...681.1183K}, \citet{2013ApJ...765..140A}, and \citet{2014ApJ...797..126S} have shown that the strengths of the correlations between stellar mass and metallicity, as well as sSFR and metallicity, depend on how the metallicities were measured and what calibration scale was used (see also \citealt{Modjaz2008}). This means that care must be taken when comparing galaxy samples with metallicities measured using different methods and scales. In this work, we only use Galspec metallicities and measure empirical galaxy scaling relations for the specific galaxies in our metallicity subsample. Thus, even though \citet{2014ApJ...797..126S} have shown that the correlation between sSFR and metallicity using the T04 method is weaker than when using metallicities measured with other methods, our conclusion that any correlation between the SN rates and one galaxy property can be converted, through galaxy scaling relations, into the measured correlations with the two other galaxy properties examined here is internally consistent.

\section{Discussion}
\label{sec:discuss}

Here, we compare the rate correlations we observe to similar measurements from the literature as well as predictions from theoretical models.

\subsection{CC SN Rates vs. Number Ratios}
\label{subsec:CCSN_context}

\begin{figure}
 \includegraphics[width=0.47\textwidth]{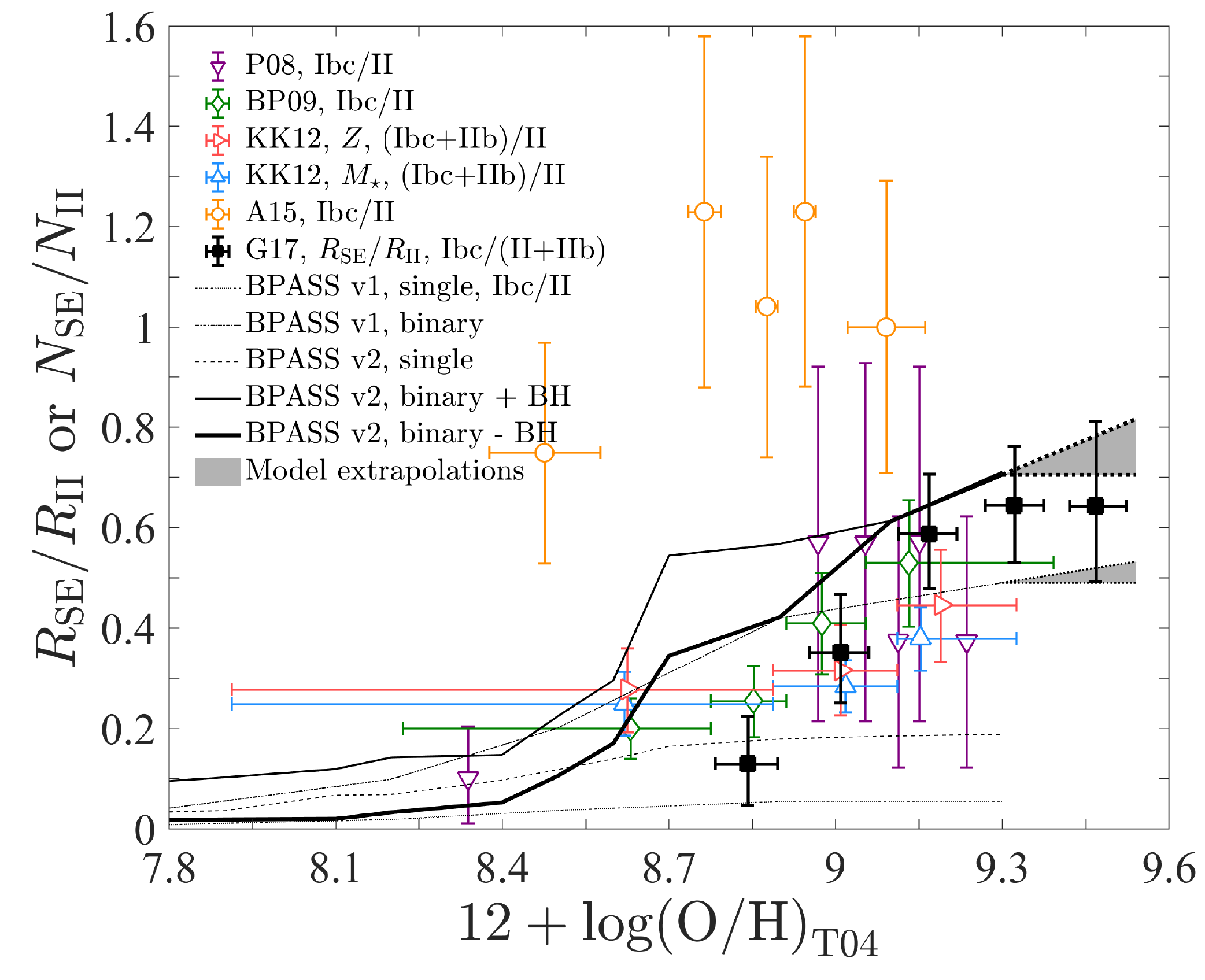}
 \caption{The ratio between the SE SN and SN II rates, as a function of metallicity (G17, i.e., this work, black squares). Previous works from the literature, which measured the ratio between the numbers of SE SNe and SNe II, $N_{\rm SE}/N_{\rm II}$, are shown as open symbols: \citet[P08, purple downturned triangles]{2008ApJ...673..999P}; \citet[BP09, green diamonds]{2009A&A...503..137B}; \citet[KK12, $Z$: red right-facing triangles, based on spectroscopic metallicities; KK12, $M_\star$: blue upturned triangles, based on stellar masses and the mass--metallicity correlation of \citealt{2004ApJ...613..898T}]{2012ApJ...759..107K}; and \citet[A15, orange circles]{2015PASA...32...19A}. Our measurements are consistent with all past studies, except for \citet{2015PASA...32...19A}, whose combined SN sample was biased towards SE SNe. We add measurements at $12+\rm{log(O/H)}\gtrsim9.2$, where the ratio between the rates levels off. BPASS models (v1: \citealt{2008MNRAS.384.1109E}, v2: \citealt{2016MNRAS.462.3302E,2016MNRAS.456..485S}) for single stars (dotted and dotted--dashed) are inconsistent with all measurements. Binary models (dashed and solid), on the other hand, are broadly consistent with the observations, whether they include black holes produced during core collapse ($+$BH, thin solid) or not ($-$BH, thick solid).}
 \label{fig:ncomp}
\end{figure}

In Figure~\ref{fig:ncomp}, we compare between our measurements of the ratio between the SE SN and SN II rates, $R_{\rm SE}/R_{\rm II}$, and measured ratio of the \emph{numbers} of SE SNe and SNe II, $N_{\rm SE}/N_{\rm II}$ from the literature, as well as model predictions for $R_{\rm SE}/R_{\rm II}$. We find that our rate ratios are broadly consistent with previous number-ratio measurements, as well as models that assume a significant fraction of SE SNe arise from interacting binary systems, rather than solely single stars. 

As we explain in detail below, our measurements differ from, and add to, previous works on the following points: (1) they are based on absolute rates, as opposed to number ratios; (2) they are derived from the well-understood, homogeneous LOSS SN sample; and (3) they sample higher metallicity values than other studies, at which the ratio might be leveling out instead of continuing to increase monotonically (a statistically significant trend, as we have shown in Section~\ref{subsec:rates_mass}).

Figure~\ref{fig:ncomp} compares between several studies, each of which included different types of SNe in their ratios and used different methods to estimate metallicities. 

\citet{2008ApJ...673..999P}, \citet{2009A&A...503..137B}, and \citet{2015PASA...32...19A} measured the number ratio of SNe Ibc to SNe II. In these cases, ``SNe Ibc'' referred to SNe Ib, Ic, and SNe that might have been either of these (usually referred to as SNe Ib/c). These studies do not mention SNe IIb by name, and we assume they were included in the SN II category (for \citealt{2015PASA...32...19A}, this has been ascertained through private communication with J. Anderson). \citet{2012ApJ...759..107K} explicitly included SNe IIb in the numerator of their number ratio (i.e., Ibc+IIb/II), whereas we included SNe IIb in the denominator, as explained in Section~\ref{subsec:rates_mass}. In that section, we also showed that moving the SNe IIb from the SN II to SE SN column did not have an appreciable effect on our rate-ratio measurements. 

As for metallicity, \citet{2008ApJ...673..999P} and \citet{2012ApJ...759..107K} used SDSS metallicities measured by the MPA-JHU pipeline, using the T04 scale (see Section~\ref{subsec:OH} for details). \citet{2012ApJ...759..107K} specifically chose SDSS fibers closest to the SN explosion sites. \citet{2009A&A...503..137B} relied on the correlation between galaxy luminosity (i.e., mass) and metallicity to estimate global metallicities, and galaxy metallicity gradients to estimate metallicities at the explosion sites (in Figure~\ref{fig:ncomp}, we reproduce the latter). \citet{2015PASA...32...19A} conducted a meta-analysis of oxygen abundances in host-galaxy HII region at the SN explosion sites, in the PP04-O3N2 scale \citep{2004MNRAS.348L..59P}. Here, we have converted these values to the T04 scale via the conversion factors from \citet{2008ApJ...681.1183K}. Finally, we converted our rate-ratio measurements as a function of stellar mass from Figure~\ref{fig:likelihood_ratio} into rate ratios vs. metallicity (reported in Table~\ref{table:rate_ratio}) via the empirical mass-metallicity correlation from Table~\ref{table:fits} (for all galaxies).

Two surveys are not represented in Figure~\ref{fig:ncomp}. \citet{2012IAUS..279...34A} measured $N_{\rm Ibc}/N_{\rm II}$ ratios (SNe IIb were treated as SNe II) as a function of host-galaxy luminosity, but there does not seem to be a significant trend in their measurements. We do not show these measurements in Figure~\ref{fig:ncomp}, but note that as they vary in the range $\approx 0.2$--$0.4$, they would be broadly consistent with the other measurements in the plot. \citet{2012A&A...544A..81H,2014MNRAS.444.2428H} used a subsample of SNe from the Asiago Supernova Catalog \citep{1999A&AS..139..531B} and the Sternberg Astronomical Institute (SAI) Supernova Catalog \citep{2004AstL...30..729T} to measure $N_{\rm Ibc}/N_{\rm II}$ in galaxies with different morphologies and disturbance levels (e.g., interacting vs. merging galaxies). They found that $N_{\rm Ibc}/N_{\rm II}$ is lower in late-type (Sc--Sm) galaxies than in early-type (S0/a--Sbc) at a 5\% significance level, which is consistent with the trend we observe. As in this work, \citet{2014MNRAS.444.2428H} included SNe~IIb in the SN II bin. 

The biggest difference between our measurements and those of previous studies is that we measure the ratio of SN \emph{rates}, as opposed to numbers. The rates take into account both the numbers of observed SNe and, through the control times, the survey's sensitivity to different types of SNe. Normalized by the mass of the stars surveyed for SNe in each galaxy, these rates are an accurate representation of the numbers of SNe produced by a given stellar population. SN numbers, on their own, can also be used to compare between different SN types and to connect between SNe and local stellar populations, but only if those numbers come from a complete sample that takes into account SNe missed by the survey.\footnote{We expand on this in Paper II, where we use a complete subsample of LOSS SNe to measure SN population fractions.} 

\citet{2012IAUS..279...34A} does not discuss the composition of the Palomar Transient Factory (PTF) sample, so its completion cannot be ascertained. \citet{2009A&A...503..137B}, \citet{2012ApJ...759..107K}, and \citet{2014MNRAS.444.2428H} used the Asiago and SAI catalogs. Both of these are inhomogeneous collections of SNe reported by different surveys, each with its own, sometimes unknown, detection and classification biases.\footnote{For example, \citet{2009A&A...503..137B} report a ratio of $1.53\pm0.35$ between SNe Ic and Ib, whereas in \citet{2016arXiv160902922S}, we show that with proper classification, the ratio is actually $0.6\pm0.3$.}

\citet{2008ApJ...673..999P} and \citet{2009A&A...503..137B} limited the redshift range of their SN samples in order to turn them into quasi-complete samples. \citet{2012ApJ...759..107K} found that the different CC SN subtypes in their sample broadly followed the same redshift distribution, from which they concluded that the surveys from which these SNe originated had similar control times for the various CC SN subtypes. The consistency of the $N_{\rm SE}/N_{\rm II}$ measurements of these studies with our $R_{\rm SE}/R_{\rm II}$ measurements shows that these attempts to make their samples complete were, on the whole, successful. 

On the other hand, \citet{2015PASA...32...19A} did not try to limit their sample. They drew SNe from several studies that attempted to measure metallicities at the SN explosion sites (see their section 3.3), and so were biased toward SE SNe. This explains why their measurements are inconsistent with all the others in Figure~\ref{fig:ncomp}, and why their number ratios are biased to more SE SNe rather than SNe II. As we measure rate ratios, and the rates are derived from the homogeneous, well-understood LOSS SN sample, our measurements are not subject to these concerns.

Because the LOSS galaxy sample is biased toward massive galaxies, our rate ratios cover a range of higher metallicity values that was not covered by previous surveys. In this range, the rate ratio might be leveling out instead of continuing to rise monotonically, as one might extrapolate from previous studies. If metallicity alone is the driver of the CC SN rate correlations, this plateau would imply that above some threshold, higher metallicity values would have no effect on SE SN progenitors.

\subsection{Constraints on SE SN progenitors}
\label{subsec:models}

In Figure~\ref{fig:ncomp}, we compare the different rate- and number-ratio measurements to model predictions derived with the Binary Population and Spectral Synthesis code (BPASS; version 1: \citealt{2008MNRAS.384.1109E}, version 2: \citealt{2016MNRAS.462.3302E,2016MNRAS.456..485S}). The BPASS models include predictions for single-star as well as interacting binary progenitors. The BPASS v2 binary models are split between a model that includes all SNe produced during core collapse and a model where SNe that produce a black hole during core collapse are removed. There is growing evidence that at least for some SNe, formation of a black hole makes it difficult to observe the events \citep{2015MNRAS.450.3289G,2015PASA...32...16S}. Therefore, removing such events from model predictions provides an estimate of their contribution to the SN rates.

While observational studies report oxygen abundances, stellar evolutionary models use the metallicity mass fraction $Z$, and in particular the iron mass fraction of the SN progenitor, since it sets the mass loss of the pre-explosion star (e.g., \citealt{2005A&A...442..587V}) and the opacity of the stellar envelope (and therefore, e.g., the lifetime and luminosity of the star on the main sequence). Given the uncertainty in the measurement for the solar oxygen abundance, as well as in the relationship between iron and oxygen abundances (for a review on SN metallicity studies and their caveats, see, e.g., \citealt{2011AN....332..434M,2015PASA...32...19A}), the systematic uncertainties in comparing metallicities from stellar evolutionary models to observed abundances are estimated to be on the order of 0.1 dex.

The BPASS models based on single-star evolution consistently fail to produce enough SE SNe (see also, e.g., \citealt{2011MNRAS.412.1522S}), as the minimum initial mass for SE SN progenitors is quite constant at high metallicities (the minimum mass for SN II progenitors rises slightly with increasing metallicity, but the overall rate is mostly affected by the minimum mass set by the initial mass function (IMF)) and because they are limited by the amount of mass loss that stars with $M_\star<20~{\rm M_\sun}$ require to strip their hydrogen layer before exploding.

Models that assume that a majority of SE SNe are produced in binaries are broadly consistent with the various number and rate ratios, whether they include SNe that produce black holes during core collapse or not. However, it is intriguing that all such models are offset from the rate ratio at $12+{\rm log(O/H)}>9$ by $\sim0.2$ dex (though this may be consistent with the systematic uncertainties in comparing metallicities from stellar evolutionary models to observed abundances). The BPASS models do not extend beyond $12+{\rm log(O/H)}=9.3$, but their extrapolation to higher metallicities (shown as the area between the last measured value and the linear extrapolation of that value) is consistent with our measurements.

These conclusions are consistent with previous studies that have preferred binaries over single stars as SE SN progenitors, e.g., through studies of relative rates (e.g., \citealt{2011MNRAS.412.1522S,2016arXiv160902922S}) or comparisons of the ejecta masses of observed SE SNe (e.g., \citealt{2011ApJ...741...97D,2013MNRAS.434.1098C,2016MNRAS.457..328L}) and the estimated masses of Wolf--Rayet stars at the time of explosion (e.g., \citealt{2003A&A...404..975M,2015PASA...32...15Y}). We note that \citet{2013A&A...558A.131G} used rapidly rotating single stars to reproduce observed SE SN rates, but had to invoke high mass-loss rates, which might not be physical (e.g., \citealt{2014ARA&A..52..487S}).

The most direct method to identify SN progenitors remains the detection of candidate progenitors in pre-explosion imaging. The majority of SE SN pre-explosion observations provide only nondetections and thus upper limits on the luminosity of the progenitors (e.g., \citealt{2012ATel.4199....1G,2013MNRAS.436..774E,2015PASA...32...16S}). The only case of a detected SE SN progenitor was for the SN Ib iPTF 13bvn by \citet{2013ApJ...775L...7C}, who identified the progenitor as a Wolf--Rayet star. \citet{2014AJ....148...68B}, on the other hand, argued that the pre-explosion source and the ejecta mass derived from the SN light curve were consistent with an interacting binary as the progenitor. This was consistent with the prediction of \citet{2012A&A...544L..11Y} that a lower-mass helium star in a binary would be the first SE SN progenitor to be detected. \citet{2013A&A...558L...1G} suggested a rapidly rotating star as an alternative explanation, but its final ejecta mass would have been higher than that inferred from the light curve. \citet{2016MNRAS.461L.117E} and \citet{2016ApJ...825L..22F} have reported the disappearance of the progenitor in late-time images and concluded that the progenitor was part of a binary system. 

In summary, for single stars to reproduce the observational constraints described above, they typically require either rapid rotation or high mass-loss rates. These are inconsistent with the rotation and mass-loss rates of observed stellar populations. Binary models, and specifically the BPASS models used in this work, are able to match all of these observation \citep{2013MNRAS.436..774E,2016MNRAS.461L.117E,2016MNRAS.457..328L}, using a distribution of binaries that is similar to the observed distribution, and without requiring fine tuning (e.g., \citealt{2012Sci...337..444S}).

\subsection{Correlation vs. Causation}
\label{subsec:rates_meaning}

Our results provide an ideal example of the old adage that ``correlation does not imply causation.'' Although we measure several strong correlations between SN rates (of various types) and different galaxy properties, we show throughout this paper that by using well-known galaxy scaling relations we can turn a correlation between the SN rates and one galaxy property into any of the other correlations measured here. This means that we cannot tell which galaxy property, or combination of properties, is the cause of these correlations. 

There are theoretical reasons to expect that the fates of massive stars could be determined in part by metallicity (e.g., \citealt{2012ARA&A..50..107L,2014ARA&A..52..487S,2015PASA...32...15Y}) or SFR conditions (e.g., because galaxies with high sSFRs may have altered IMFs or because they may possess dense clusters with a high number of dynamical interactions; \citealt{2010ApJ...717..342H,2011MNRAS.415.1647G,2012Sci...337..444S,2013ApJ...771...29G,2013MNRAS.436.3309W,2014MNRAS.442.1003P}), but it is not clear how the total mass of the galaxy would impact the death of one or two massive local stars. On their own, though, our data do not allow us to either prove or disprove these claims.

It is tempting to assume that SN~II rates are only dependent on a galaxy's SFR, or that the SE SN rates depend on metallicity. Previous works have made such claims (e.g., \citealt{2009A&A...503..137B}). We made a similar error when, in G15, we claimed that our model for the SN~Ia rates was self-consistent because it fit not only the rates vs. stellar mass, but also the rates vs. SFR and sSFR; because of the galaxy scaling relations, once a model is fit to one correlation it will automatically fit the others.

Thus, the mere existence of correlations between SN rates and galaxy properties cannot be used to constrain SN progenitor models. Instead, we suggest concentrating on emergent structures within the SN rate correlations. For example, the model we use to explain the SN~Ia rate correlations predicts that the rates should plateau in galaxies with $M_\star<10^9~{\rm M}_\sun$ and $M_\star>10^{11}~{\rm M}_\sun$. Likewise, any model for the progenitors of SE~SNe, whether it depends on metallicity, binarity, or rotation (or some combination of these properties; e.g., \citealt{2013ApJ...764..166D,2014ApJ...782....7D}), should explain why the efficiency of SE~SN production, relative to SN~II production, rises as a function of galaxy stellar mass, but levels out in galaxies more massive than $\sim 2 \times 10^{10}~{\rm M}_\sun$. Such a model should also produce a smooth dependence between the rates and the different galaxy properties examined here. For example, \citet{2013ApJ...765L..43I} predict a complicated relation between the metallicity of CC~SN progenitors and the minimum mass at which they should explode. Within the metallicity range tested here, this relation is smooth and broadly consistent with our measurements. However, if the SN~II and SE~SN rates remain smooth over a larger metallicity range, that would pose a challenge to this model.

Because the galaxy scaling relations connect between the different SN correlations shown here, we also suggest concentrating on measurements of the rates as a function of either galaxy stellar mass or luminosity, as these are the most straightforward properties to measure. 

\begin{figure}
 \includegraphics[width=0.47\textwidth]{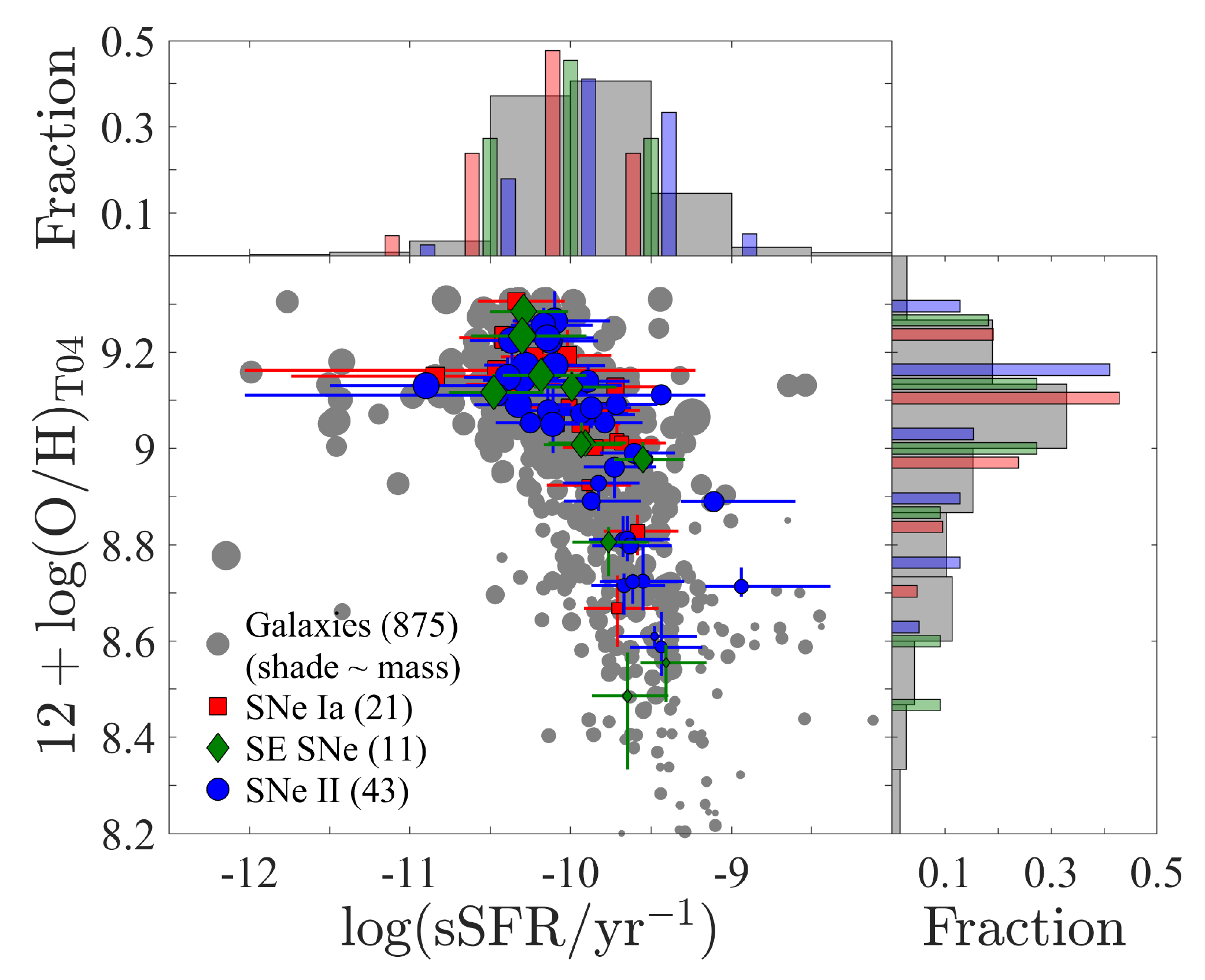}
 \caption{Specific SFR vs. metallicity of a subsample of the LOSS galaxy sample (gray circles. Host galaxies of SNe Ia, SE SNe, and SNe II are marked as in previous figures. The size of the symbols scales as ${\rm log}(M_\star/{\rm M_\sun})$ of the galaxies. The histograms on the top and side panels show the distributions of galaxies in sSFR and metallicity, respectively. The host galaxies of all SN types seem to be distributed evenly within the LOSS sample, with no preference for either sSFR or metallicity.}
 \label{fig:sSFR_OH}
\end{figure}

A possible way to ascertain which of the galaxy properties examined here is responsible for the deficiency of the SE SN rates in low-mass galaxies is to follow \citet{2016ApJ...830...13P} and \citet{2016arXiv160504925C}, who compared the stellar masses, sSFRs, and metallicities of the host galaxies of superluminous SNe with those of a complete local galaxy sample (from the Local Volume Legacy Survey, which includes 258 galaxies out to 11 Mpc; \citealt{2008ApJS..178..247K,2009ApJ...703..517D}). Such a comparison allows one to test whether the SN host galaxies diverge from the rest of the galaxies in one of the galaxy parameters, or in two or more. Although the LOSS galaxy sample is not complete, we can still use it to test for divergences within the sample.

Figure~\ref{fig:sSFR_OH} shows a subsample of 875 LOSS galaxies that have both Galspec metallicities and sSFR measurements derived from NSA photometry. Within this subsample, we find no significant divergence between the SN host galaxies and the majority of the galaxies in the subsample either in metallicity or in sSFR. We caution that this may simply reflect the size of the subsample used here ($<10\%$ of the full LOSS sample) or the lack of low-luminosity galaxies in LOSS. It would be a worthwhile endeavor to measure sSFRs and metallicities for all of the LOSS SN host galaxies, or at least for the SN-complete, volume-limited subsample (see Paper II), to mitigate the effect of the sample size. To facilitate similar tests to that in \citet{2016ApJ...830...13P}, future SN surveys should also strive to target galaxy samples that are representative of the galaxy luminosity function.

Alternatively, one could try to remove the effect of the galaxy scaling relations on the rate correlations. With a sufficiently large SN sample, such as the one that will be created by the Large Synoptic Survey Telescope \citep{2009arXiv0912.0201L}, one could single out SN host galaxies within a narrow mass range and then look for correlations with other galaxy properties. A different path would be to correlate the SN rates with explosion-site properties (the galaxy scaling relations invoked throughout this work have only been established for global galaxy properties). Integral-field unit spectroscopy of the LOSS galaxy sample would turn it into a survey of distinct star-forming regions. One could then not only measure rates as a function of local properties, but also sample the DTDs of the different SN subtypes directly, as done by \citet{Maoz2010magellan} and \citet{Maoz2010loss}.


\section{Summary and Conclusions}
\label{sec:summary}

This is the first of a series of papers in which we reanalyze the LOSS SN rates. Here, we matched the LOSS galaxy sample with SDSS and then remeasured the LOSS SN rates as a function of various global properties of galaxies: stellar mass, star-formation rates, and metallicity (in the form of nebular oxygen abundance). All of these measurements, including the control times necessary to compute them, are made public through the various tables in this work. We make the following observations.
\begin{enumerate}
 \item The specific SN II rates are strongly correlated with all galaxy properties measured here. SE SN and SN Ia rates show strong correlations with stellar mass and sSFR, but not with nuclear metallicity.
 \item The SN Ia rate--mass correlation is statistically significant in star-forming galaxies, but not in passive ones, as also noted by \citet{2006ApJ...648..868S}.
 \item The SN Ia rates are well fit by a model that combines a $t^{-1}$ DTD with the galaxy mass--age scaling relation (Figures~\ref{fig:sSFR_Ia} and \ref{fig:OH_Ia}), as suggested by \citet{Kistler2011}, \citet{GraurMaoz2013}, and G15.
 \item The ratio between SE SN and SN II rates rises with galaxy stellar mass until it flattens in galaxies with $M_\star \ga 10^{10}~{\rm M}_\sun$ (or $12+{\rm log(O/H)}\ga9.2$) (Figures~\ref{fig:likelihood_ratio} and \ref{fig:ncomp}). This trend is statistically significant, at a $>3\sigma$ level.
 \item The rate ratio measurements rule out single stars as progenitors of SE SNe, but are consistent with models that assume binary-system progenitors, as suggested by earlier works (Figure~\ref{fig:ncomp}).
 \item Similar deficiencies in the SE SN rates relative to the SN~II rates are seen when correlating the rates with other galaxy properties, though those trends are not statistically significant (Figures~\ref{fig:rates_CC_sSFR} and \ref{fig:rates_OH}). 
 \item The SN~II and SE~SN correlations do not exhibit significant breaks, which means that their underlying dependence on any of the galaxy properties studied here (or a combination of these properties) must be smooth within the dynamical range probed in this work.
 \item SE~SN host galaxies follow the same distribution in sSFR vs. metallicity space as SN~II host galaxies and LOSS galaxies that did not host SNe during the survey.
\end{enumerate}

Although the correlations shown here are broadly consistent with those shown in previous studies, the results of this work differ from previous studies by being based on absolute SN rates derived from a homogeneous, well-characterized SN sample. The LOSS sample, which is biased toward massive galaxies, has allowed us to sample a higher metallicity range than previous studies. Interestingly, in this range, the statistically significant correlation between the ratio of SE~SN to SN~II rates and galaxy stellar mass (or metallicity) levels off instead of continuing to increase monotonically.

We have shown that, owing to the known galaxy scaling relations, any correlation between the SN rates---of any SN type---and a specific galaxy property can be transformed into the measured correlations with the other galaxy properties studied here. This precludes us from ascertaining which of the galaxy properties (or some combination of them) is responsible for the correlations we observe or for the deficiency of SE~SNe in low-mass galaxies. We have outlined several methods that might allow us to bypass this problem in future experiments.

Finally, we have also enriched the LOSS sample with additional galaxy properties and the publication of the SN control times, so that further studies can be undertaken with this sample.


\section*{Acknowledgments}

We thank the anonymous referee for helpful suggestions and comments. We further thank Iair Arcavi, Ryan Chornock, Jenny Greene, Patrick Kelly, Dan Maoz, Asaf Pe'er, Michael Shara, Nathan Smith, and Todd Thompson for their comments on this work. 

O.G. is supported in part by National Science Foundation (NSF) award AST-1413260 and by an NSF Astronomy and Astrophysics Fellowship under award AST-1602595. F.B.B. is supported in part by the NYU/CCPP James Arthur Postdoctoral Fellowship. M.M. is supported in parts by NSF CAREER award AST-1352405 and by NSF award AST-1413260. A.V.F. and the Lick Observatory Supernova Search (LOSS) have received generous financial assistance from the TABASGO Foundation, US Department of Energy (DoE) SciDAC grant DE-FC02-06ER41453, DoE grant DE-FG02-08ER41653, and many NSF grants (most recently AST-1211916). KAIT and its ongoing operation were made possible by donations from Sun Microsystems, Inc., the Hewlett-Packard Company, AutoScope Corporation, Lick Observatory, the NSF, the University of California, the Sylvia \& Jim Katzman Foundation, and the TABASGO Foundation. Research at Lick Observatory is partially supported by a generous gift from Google. We thank the Lick staff for their assistance at the observatory. J.J.E. acknowledges support from the University of Auckland. The BPASS models values were calculated thanks to the contribution of the NeSI high performance computing facilities and the staff at the Centre for eResearch at the University of Auckland.

This research has made use of NASA's Astrophysics Data System and the NASA/IPAC Extragalactic Database (NED) which is operated by the Jet Propulsion Laboratory, California Institute of Technology, under contract with NASA. We also acknowledge the usage of the HyperLeda database (\url{http://leda.univ-lyon1.fr}).

Funding for the SDSS and SDSS-II has been provided by the Alfred P. Sloan Foundation, the Participating Institutions, the National Science Foundation, the U.S. Department of Energy, the National Aeronautics and Space Administration, the Japanese Monbukagakusho, the Max Planck Society, and the Higher Education Funding Council for England. The SDSS Web Site is \url{http://www.sdss.org/}. The SDSS is managed by the Astrophysical Research Consortium for the Participating Institutions. The Participating Institutions are the American Museum of Natural History, Astrophysical Institute Potsdam, University of Basel, University of Cambridge, Case Western Reserve University, University of Chicago, Drexel University, Fermilab, the Institute for Advanced Study, the Japan Participation Group, Johns Hopkins University, the Joint Institute for Nuclear Astrophysics, the Kavli Institute for Particle Astrophysics and Cosmology, the Korean Scientist Group, the Chinese Academy of Sciences (LAMOST), Los Alamos National Laboratory, the Max-Planck-Institute for Astronomy (MPIA), the Max-Planck-Institute for Astrophysics (MPA), New Mexico State University, Ohio State University, University of Pittsburgh, University of Portsmouth, Princeton University, the United States Naval Observatory, and the University of Washington.

The NASA-Sloan Atlas was created by Michael Blanton, with extensive help and testing from Eyal Kazin, Guangtun Zhu, Adrian Price-Whelan, John Moustakas, Demitri Muna, Renbin Yan and Benjamin Weaver. Renbin Yan provided the detailed spectroscopic measurements for each SDSS spectrum. David Schiminovich kindle provided the input GALEX images. We thank Nikhil Padmanabhan, David Hogg, Doug Finkbeiner and David Schlegel for their work on SDSS image infrastructure. Funding for the NASA-Sloan Atlas has been provided by the NASA Astrophysics Data Analysis Program (08-ADP08-0072) and the NSF (AST-1211644).


\software{AstroML \citep{astroMLText}, MATLAB, MPA-JHU Galspec pipeline \citep{2003MNRAS.341...33K,2004MNRAS.351.1151B,2004ApJ...613..898T}, WebPlotDigitizer}


\appendix

\section{Sample Construction Details}
\label{appendix:samples}

L11 measured the LOSS SN rates in a subsample of the LOSS galaxies, termed the ``full-optimal'' sample, which excluded (1) highly inclined galaxies, namely those galaxies with inclinations $i>75^\circ$, where the inclinations were calculated according to the formula from \citet{1926ApJ....64..321H} and measurements of the apparent major and minor axes of the galaxies (see Equation 2 of \citealt{2011MNRAS.412.1419L}); and (2) small (major axis $< 1^\prime$), early-type (E--S0) galaxies, as those were found to have lower SN detection efficiencies, at any SN magnitude, relative to the other galaxies in the sample, due to SNe being obscured by the bright nuclear regions of these galaxies. L11 further restricted the sample used for rate measurements to galaxies of Hubble types E--Scd for SNe Ia and Sab--Scd for CC SNe, as the other types of galaxies hosted very few ($<5$) or no SNe during the survey. We also use the full-optimal subsample, but do not apply this last criterion to the LOSS galaxy sample.

To estimate the masses of the LOSS galaxies, \citet{2011MNRAS.412.1419L} used $B$- and $K$-band photometry acquired from the HyperLeda database and Equation 1 from \citet{2005A&A...433..807M}, reproduced here as
\begin{equation}\label{eq:mass}
 {\rm log} \left( \frac{M_\star/L_K}{{\rm M_\odot/L_\odot}} \right) = 0.212 (B-K) -0.959,
\end{equation}
where $M_\star$ is the stellar mass of the galaxy and $L_K$ is its luminosity in the $K$ band.

\begin{figure*}
 \centering
 \begin{tabular}{ccc}
  \includegraphics[width=0.31\textwidth]{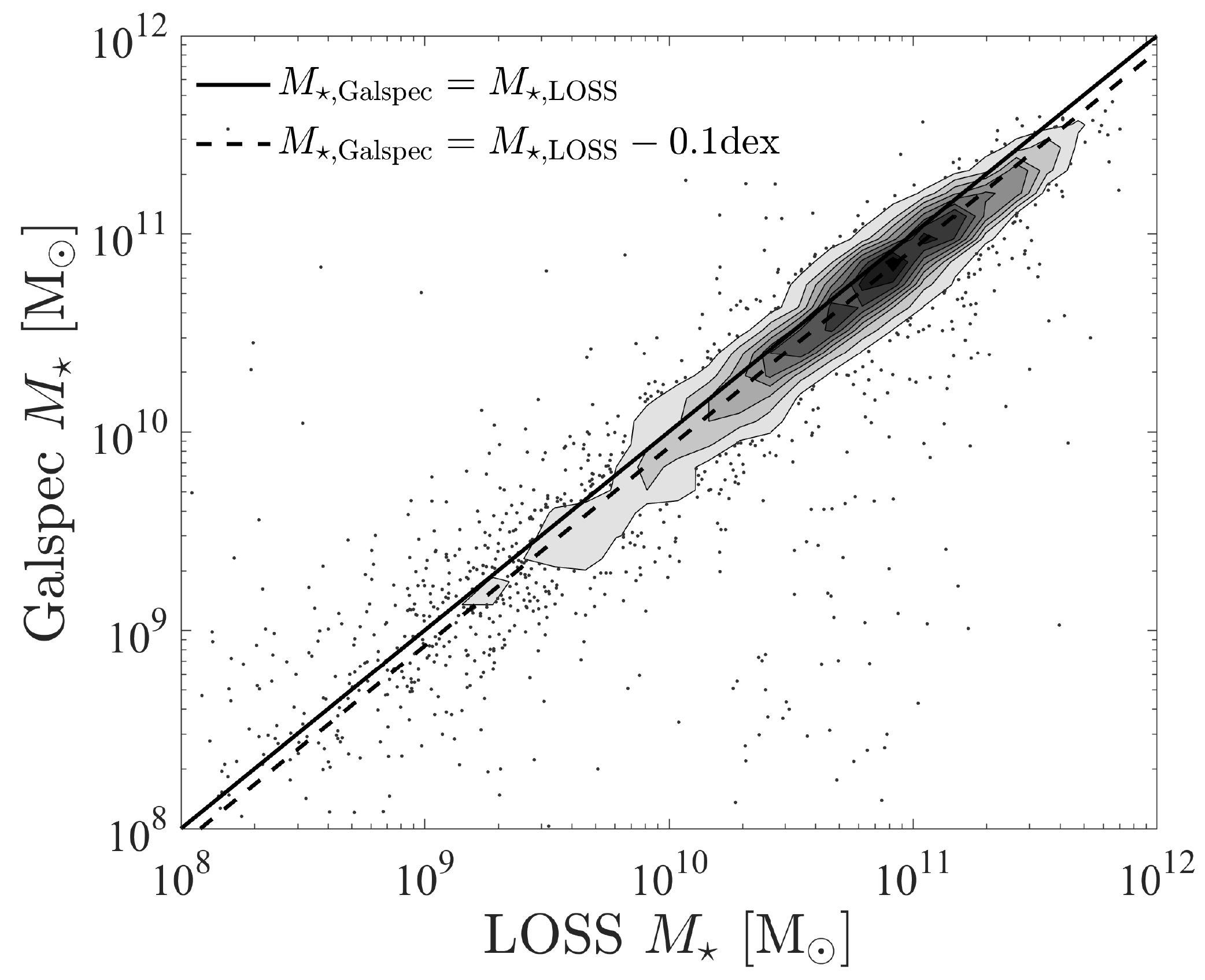} &
  \includegraphics[width=0.31\textwidth]{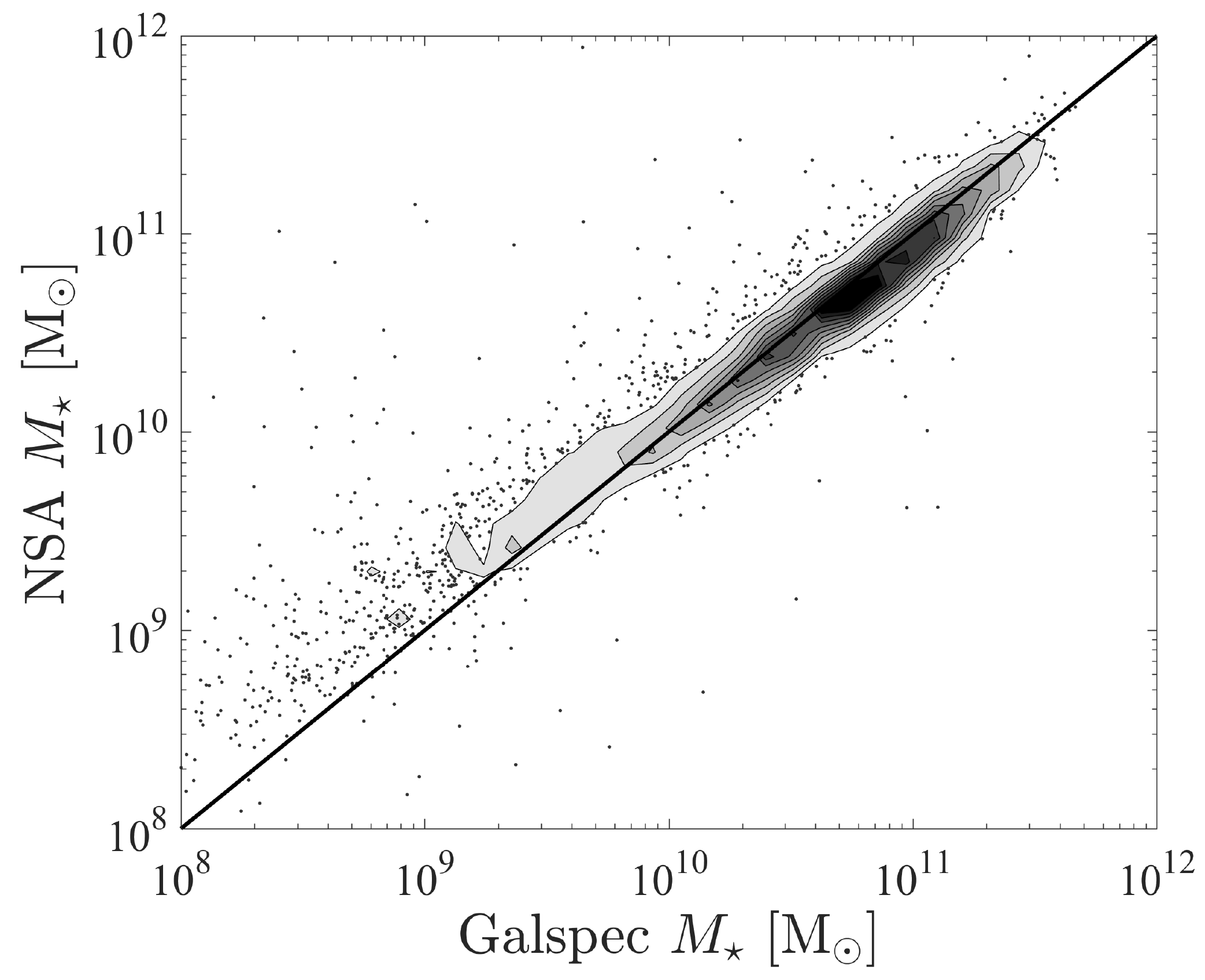} &
  \includegraphics[width=0.31\textwidth]{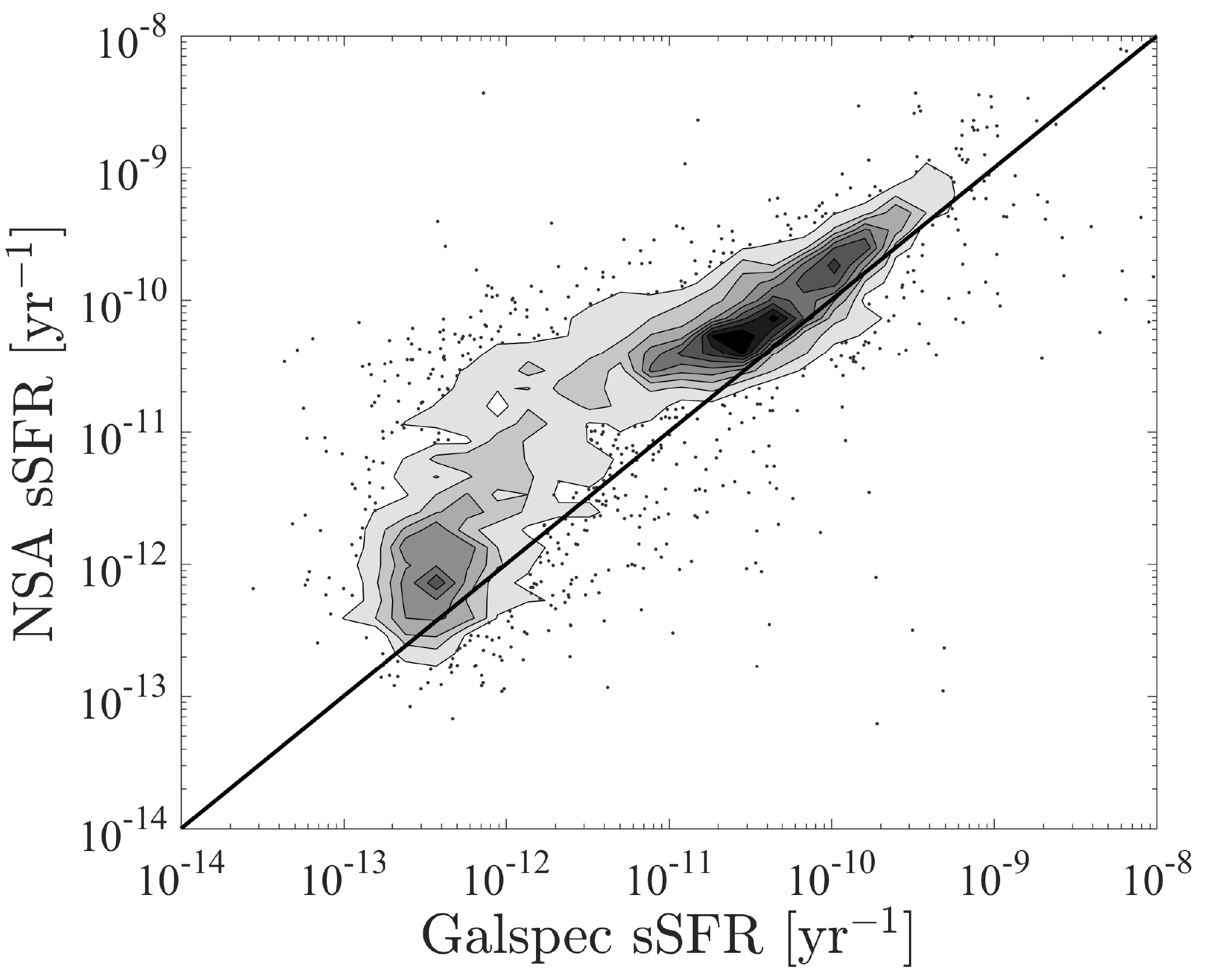}
 \end{tabular} 
 \caption{Left: A comparison between the Galspec and LOSS stellar masses reveals that, though overall consistent with each other, the latter are larger by a factor of $1.2$ than the former. Center: Stellar masses estimated from NSA $ugriz$ Petrosian photometry are consistent with Galspec masses. Right: sSFRs, estimated from NSA $ugriz$ Sersic photometry, are higher than Galspec sSFRs, which suffer from galaxy shredding. In all three panels, the solid line represents 1:1 equivalence.}
 \label{fig:mass}
\end{figure*}

In Figure~\ref{fig:mass}, we compare the stellar masses of 3855 galaxies that have nonzero LOSS and Galspec stellar masses, and find that the LOSS masses are systematically larger than the Galspec masses by 0.1 dex. For consistency between the LOSS and SDSS ``sSFR'' and ``metallicity'' samples, throughout this work we scale down the LOSS stellar masses by a factor of $1.2$. This difference may be attributed to the different IMFs used by each method: a ``diet'' \citet{1955ApJ...121..161S} IMF for the LOSS masses \citep{2005A&A...433..807M}, as opposed to a \citet{2001MNRAS.322..231K} IMF for the Galspec masses \citep{2007ApJS..173..267S}. It is important to note that, overall, the Galspec masses are consistent with the LOSS masses. This empirical test shows that the Galspec masses do not suffer the same bias as the SFRs, as described in Section~\ref{subsec:SFR}.

Of the 10121 galaxies in the LOSS sample, there are 866 galaxies (8.6\%) without known masses owing to a lack of either $B$- or $K$-band photometry. L11 chose not to use these galaxies when measuring their SN rates. Here, we correlate between the existing stellar masses and luminosities in order to interpolate the missing stellar masses. We divide the galaxies into bins of width 0.2 dex in either $L_B$ or $L_K$. For a specific galaxy with unknown mass, we assign the median of the masses within its luminosity bin as its mass value, and take the 16th and 84th percentiles of the mass distribution within the bin as the mass values' uncertainty. In Paper II, we show an example of this procedure for a subsample of the LOSS galaxies. Of the 866 galaxies with missing masses, four are outliers with low luminosities, which result in near-zero masses. These galaxies, which do not host any SNe, are excluded from the final sample.

In this work, we use two SDSS value-added catalogs---the MPA-JHU Galspec galaxy properties and the NASA-Sloan Atlas (NSA) photometry---to measure specific SN rates as a function of sSFR and metallicity. We refer the reader to \citet{2003MNRAS.341...33K}, \citet{2004MNRAS.351.1151B}, and \citet{2004ApJ...613..898T} for a thorough description of the Galspec pipeline. We initially used the Galspec measurements to construct the ``sSFR'' sample and ``metallicity'' samples. We first cross-matched the coordinates of the LOSS galaxies, as given in Table 2 of \citet{2011MNRAS.412.1419L}, with the SDSS coordinates of all the galaxies analyzed with Galspec, requiring that any two sets of coordinates be no more than $3\arcsec$ (the diameter of the SDSS fiber aperture) apart. Of the 14,882 LOSS galaxies, 4196 ($\sim28\%$) were matched with SDSS galaxy spectra. For the SDSS sSFR sample, we select only those that have nonzero stellar mass, SFR, and sSFR values (4040 galaxies), and are part of the LOSS full-optimal sample. For the SDSS metallicity sample, we also require nonzero metallicity values. As these values were only measured for those SDSS galaxies classified as ``star-forming'' \citep{2004ApJ...613..898T}, this subsample (1000 galaxies) represents a subset of the galaxies in the SDSS sample.

The host galaxies of CC SNe tend to be star-forming. However, the SDSS pipeline photometry suffers from shredding if multiple star-forming sites are resolved in the disks. Usually, the LOSS host galaxy is cross-matched to the central source, so that only the redder light from the galaxy bulge is picked up. The galaxy properties, especially the sSFRs, derived from such standard SDSS photometry, may not represent those of the entire galaxy or at the SN site with active star formation. To better characterize the global properties of the SDSS galaxies, we make use of NSA photometry, which improves on the original SDSS photometric analysis using the detection and deblending technique described by \citet{2011AJ....142...31B}.

To derive global stellar masses and sSFRs, we apply spectral energy distribution (SED) fitting to the NSA photometry in the five SDSS bands, adopting the methodology of \citet{2007ApJS..173..267S}. The full details of the SED fitting technique used here can be found in \citet{2012ApJ...756..113H,2012AJ....143..133H}. A library of model SEDs are generated, using the stellar population synthesis code of \citet{2003MNRAS.344.1000B}, with an extensive range of internal extinction, metallicity, and star-formation history considered. The final physical properties, including stellar mass and sSFRs, are computed as the median of all model values, where each model is weighted according to its fit likelihood.

The SDSS standard pipeline magnitudes are expected to miss the blue light from star forming regions in disk regions. As shown in the right panel of Figure~\ref{fig:mass}, we have confirmed that the NSA Sersic fluxes yield overall bluer colors and thus higher sSFRs from SED fitting, relative to those from the SDSS pipeline magnitudes. However, as shown in the central panels of Figure~\ref{fig:mass}, the stellar mass estimates, based on NSA Petrosian fluxes, are less affected by shredding, because the red central bulge dominates the mass. We use the same library of model SEDs as the MPA-JHU Galspec pipeline. As a result, our stellar mass estimates are consistent with the Galspec values for the sources with good SDSS pipeline photometry. Thus, for the sSFR sample, we adopt the stellar masses and sSFR values computed here from the NSA photometry. For the metallicity sample, however, we use the Galspec stellar mass estimates. Our estimates of the stellar masses and sSFRs, based on NSA photometry, are included in Table~\ref{table:gal_prop}.

\section{Consistency among Galaxy Samples}
\label{appendix:consistency}

\begin{figure*}
 \center
 \begin{tabular}{cc}
  \includegraphics[width=0.47\textwidth]{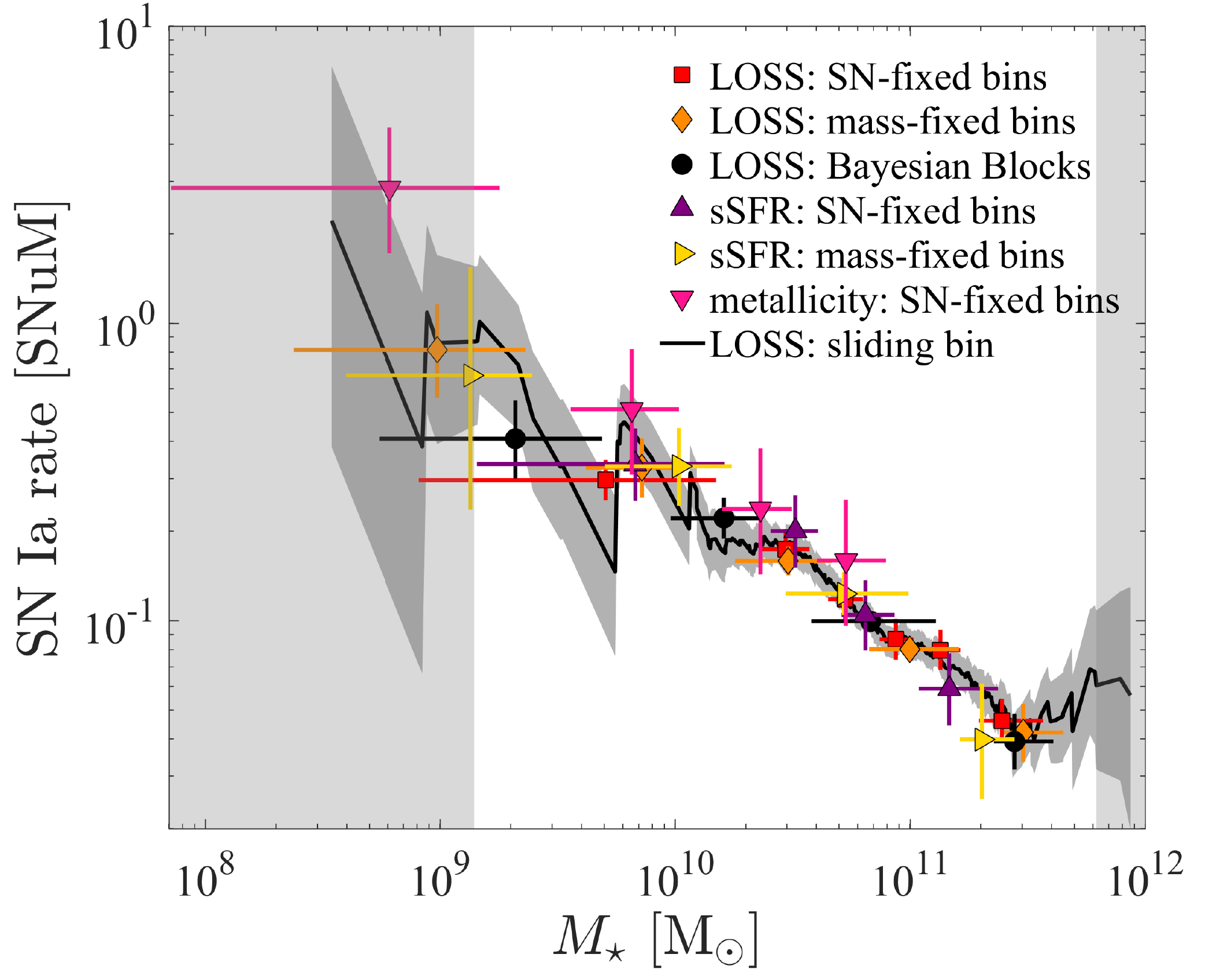} &
  \includegraphics[width=0.47\textwidth]{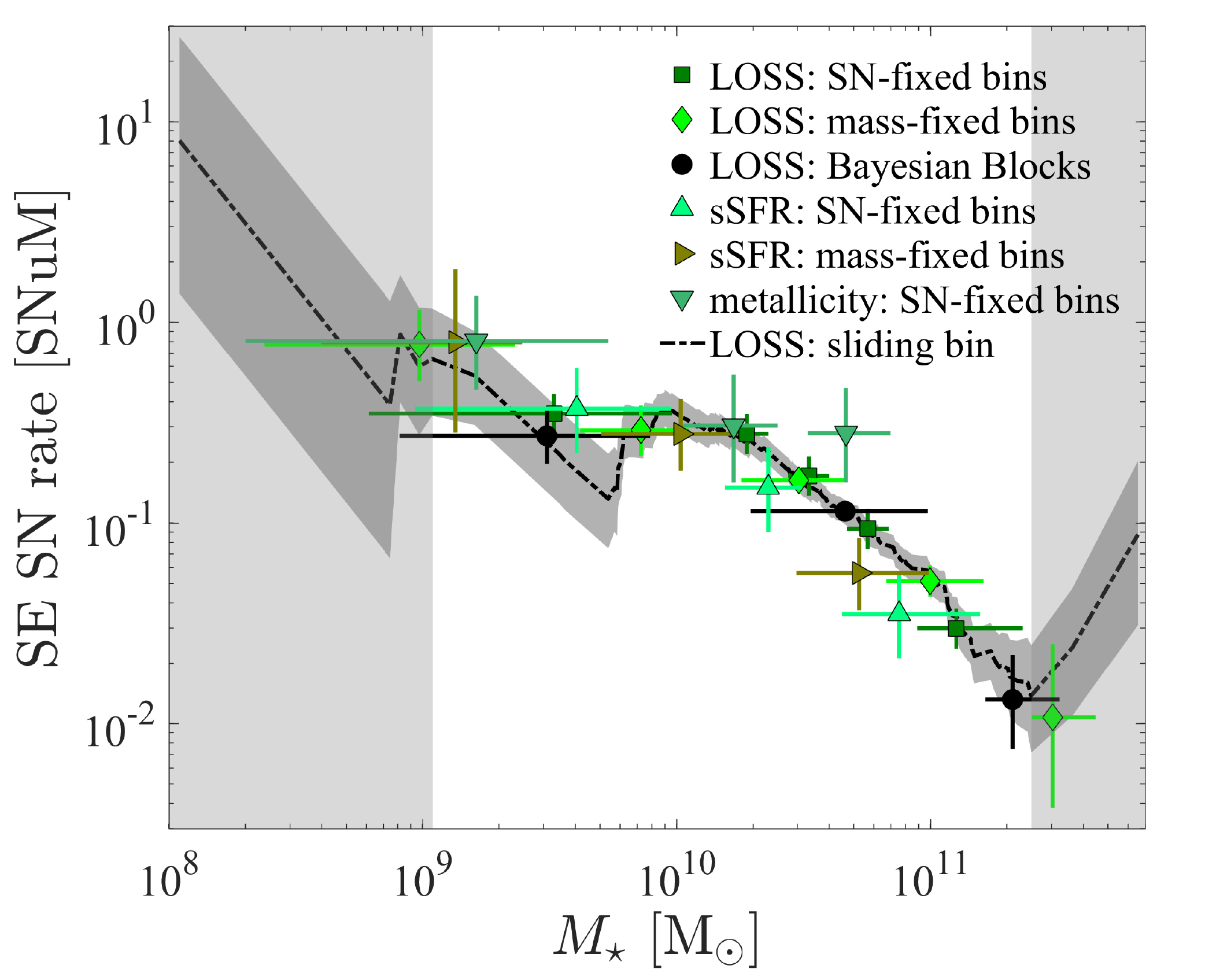} \\
  \multicolumn{2}{c}{\includegraphics[width=0.47\textwidth]{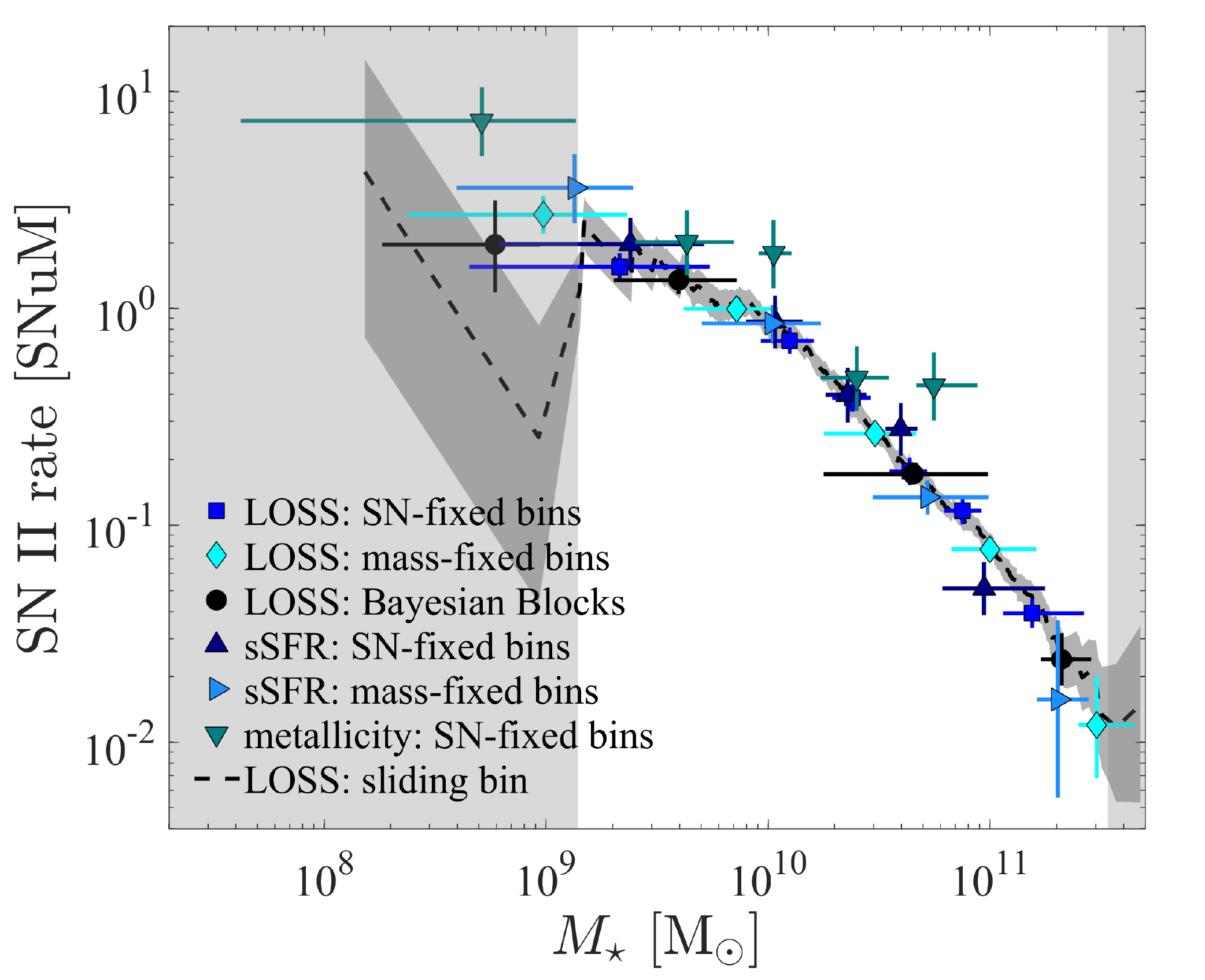}} 
 \end{tabular}
 \caption{Specific SN rates as a function of galaxy stellar mass for SNe Ia (upper left), SE SNe (upper right), and SNe II (upper center). Symbols show the rates as measured with the LOSS, sSFR, and metallicity samples, in bins with either roughly equal numbers of SNe (``SN-fixed'') or equal galaxy mass (``mass-fixed,'' in log space). The black symbols denote rates as measured in bins calculated with the Bayesian Blocks algorithm. The curves show rates measured with a sliding bin; the 68\% Poisson uncertainties of these measurements are shown as the gray regions. Light-gray patches show where the sliding-bin rates are based on $\le 3$ SNe per bin, leading to large Poisson uncertainties. The various measurements, from different samples and with different bins, are generally consistent. Note that the metallicity-sample rates are constrained to a narrower range of galaxy stellar masses and are slightly enhanced in more massive galaxies.}
 \label{fig:all_samples}
\end{figure*}

In this work, we measure specific SN rates using the LOSS ``full-optimal'' sample and two subsamples of this sample, labeled ``sSFR'' and ``metallicity,'' which are described in Section~\ref{sec:galaxies}. As one chooses progressively smaller subsamples of a given sample, there arises the possibility that any resulting measurements from those subsamples would be biased, relative to measurements performed with the main sample. To test for any such biases, and whether the sliding-bin rates are a good representation of the rates, Figure~\ref{fig:all_samples} shows the rates for each SN type as measured from the different subsamples and with various binning schemes. ``SN-fixed'' bins are chosen so that they contain roughly the same number of SNe in each bin while ``mass-fixed'' bins contain roughly the same amount of stellar mass, in log scale. Finally, we also show the rates in fixed bins of varying width, as calculated with the AstroML\footnote{\url{http://www.astroml.org/}} \citep{astroMLText} realization of the Bayesian Blocks algorithm of \citet{2013ApJ...764..167S}.

The various binning schemes produce measurements that generally agree with each other, except at the lower end of the mass range, where the small number of galaxies and SNe can cause relatively large fluctuations in the rates. It is also clear that the rates as measured with a sliding bin are a good representation of the data. Moreover, the rates measured with the sSFR subsample are consistent with those measured with the main LOSS sample. The rates measured from the metallicity subsample, however, are markedly higher in the largest-mass bin for both SNe II and SE SNe; no such bias is noticeable for the SN Ia rates. We ascribe this bias to the small size of the metallicity subsample ($\sim10$\% of the main LOSS sample). This bias means that a larger sample is required to test the validity of the correlations between the SN rates and metallicity shown in Figure~\ref{fig:rates_OH}. The connection between these correlations and the galaxy scaling relations between metallicity and stellar mass, however, should not be affected by this bias, as in this work we measure the latter scaling relation directly from the galaxies and SN hosts in the metallicity subsample.

\section{Likelihood Ratio Calculation}
\label{appendix:ratio}

The likelihood ratio test is used to compare the goodness-of-fit of nested models, such as polynomials of increasing order. In this work, we use the likelihood ratio test to compare between zeroth-, first-, and second-order polynomials as fits to various datasets. The likelihood ratio ($R_L$) is simply the ratio of the likelihoods ($L$) of the data ($x$) given the best-fitting parameters of each type of fit, $\theta_0$ for the null hypothesis (the lower-order polynomial) and $\theta_1$ for the model being tested:
\begin{equation}\label{eq:ratio}
 R_L = \frac{L(\theta_0|x)}{L(\theta_1|x)}.
\end{equation}
The likelihood ratio is distributed as a $\chi^2$ distribution, with the number of degrees of freedom equal to the difference in the degrees of freedom of the models tested (one or two in our case). $\chi^2=-2\,{\rm ln}(R_L)$, so that through the likelihood ratio we can obtain a $p$-value for the significance of the rejection of the null hypothesis. These $p$-values can then be translated into Gaussian standard deviations, so that here we report the significance of the tests in multiples of $\sigma$, with $3\sigma$ as the minimal significance for a ``discovery.''

When fitting the SN rates, the likelihood function is simply that of the Poisson probability density function (PDF), as the uncertainties of the rates are dominated by the sizes of the SN samples. The likelihood function is then
\begin{equation}\label{eq:poisson}
 L=\prod_{i=1}^N P(n_i | \lambda_i),
\end{equation}
where $N$ is the number of bins in which the rates are measured, $\lambda$ is the observed number of SNe, and $n$ is the number of SNe resulting from the best fit to the rates.

The likelihood function of the rate-ratio measurements shown in Figure~\ref{fig:likelihood_ratio} is more complicated. Formally, it should be the ratio ($w$) of the Poisson PDFs of the SE SNe, $P(x)=(\lambda_x / x!)e^{-x}$, and SNe II, $P(y)$, in each bin. However, as \citet{griffin1992distribution} notes, if $P(y)=0$, the denominator vanishes, and $R_L$ is undefined. \citet{griffin1992distribution} uses a truncated version of $P(y)$, so that $p[P(y)\leq 1]=1$, to solve this problem, but the resultant PDF of $w$ is hard to compute. However, for sufficiently large values of $\lambda_x$ and $\lambda_y$, the Poisson PDFs will approach the Gaussian PDFs $G(x)$ and $G(y)$, with means $\mu_x=\lambda_x$ and $\mu_y=\lambda_y$, and standard deviations $\sigma_x=\sqrt{\lambda_x}$ and $\sigma_y=\sqrt{\lambda_y}$. \citet{HINKLEY01121969} calculated the ratio of Normal PDFs. For Gaussian functions, \citet{doi:10.1287/mnsc.21.11.1338} have shown that given that $x$ and $y$ are uncorrelated, and that the coefficients of variation satisfy ${\rm CV}(x)>0.005$ and ${\rm CV}(y)<0.39$, the ratio $w$ can be transformed via the Geary-Hinkley transformation into
\begin{equation}
 t = \frac{\mu_y w - \mu_x}{\sqrt{\sigma_y^2 w^2 - 2\rho \sigma_x \sigma_y w + \sigma_x^2}},
\end{equation}
and the PDF of $t$ will then be a Normal distribution with $\mu=0$ and $\sigma=1$. For $G(x)$ and $G(y)$, the conditions on the coefficients are satisfied for $\lambda_x\lesssim39,200$ and $\lambda_y>7$.

The latter conditions are satisfied for all binning methods of the rate ratio. However, owing to the small SN samples, the basic condition for the approximations described above---that the numbers of SNe are large enough that their Poisson PDFs approach Gaussian ones---is clearly not satisfied. To alleviate this problem, when fitting the polynomials to the measurements, we take the upper uncertainty (which, for a Poisson distribution, is always larger than the lower uncertainty) of each measurement as the overall uncertainty.



\begin{deluxetable*}{lCCCC}
 \tablecaption{Galaxy and SN Samples.\label{table:sample_general}}
 \tablehead{
 \colhead{Sample} & \colhead{$N_{\rm Gal}$} & \colhead{$N_{\rm Ia}$} & \colhead{$N_{\rm SE}$} & \colhead{$N_{\rm II}$}}
 \colnumbers
 \startdata
  LOSS        & 10117 & 274  & 116  & 324 \\
  sSFR        & 2415  & 65   & 18   & 79  \\
  metallicity & 1000  & 24   & 14   & 52  \\
 \enddata
\end{deluxetable*}

\begin{splitdeluxetable*}{lCCCCCCCCCCCcBCCCCCCCCCCCC}
 \tablecaption{Galaxy Properties and SN Control Times.\tablenotemark{a} \label{table:gal_prop}}
 \tablehead{
 \colhead{Galaxy} & \colhead{fo} & \colhead{$N_{\rm Ia}$} & \colhead{$N_{\rm Ib/c}$} & \colhead{$N_{\rm II}$} & \colhead{$t_{c,\rm Ia}$} & \colhead{$t_{c,\rm Ib/c}$} & \colhead{$t_{c,\rm II}$} & \colhead{$M_{\star,\rm LOSS}$} & \colhead{$L_B$} & \colhead{$L_K$} & \colhead{$d$} & \colhead{$T$} & \colhead{$\alpha$} & \colhead{$\delta$} & \colhead{Plate} & \colhead{MJD} & \colhead{Fiber} & \colhead{$z$} & \colhead{$M_{\star,\rm MPA}$} & \colhead{SFR} & \colhead{sSFR$_{\rm MPA}$} & \colhead{Metallicity} & \colhead{$M_{\star,\rm NSA}$} & \colhead{sSFR$_{\rm NSA}$}
 }
 \colnumbers
 \startdata
 UGCA\_017 & 0 & 1 & 0 & 0 & 8.222 & 7.817 & 7.799 & 0.2599 & 1.4566 & 1.0383 & 25.78 & Sc & 21.5601 & -6.0942 & \cdots & \cdots & \cdots & \cdots & \cdots & \cdots & \cdots & \cdots & \cdots & \cdots \\
 UGCA\_024 & 1 & 0 & 0 & 0 & 9.772 & 9.534 & 9.599 & 0 & 0.1499 & 0 & 17.44 & Scd & 31.1308 & -6.1989 & \cdots & \cdots & \cdots & \cdots & \cdots & \cdots & \cdots & \cdots & \cdots & \cdots \\
 UGC\_03825 & 1 & 0 & 0 & 1 & 6.776 & 2.215 & 4.993 & 10.0454 & 3.3206 & 15.1121 & 115.29 & Sbc & 110.8882 & 41.4350 & 1864 & 53313 & 171 & 0.0276 & 6.7895 & 0.1012 & 0.0137 &  & 4.3351 & 0.2624 \\
 UGC\_03944 & 0 & 0 & 0 & 0 & 8.730 & 7.540 & 6.696 & 0.9475 & 1.2143 & 2.2752 & 55.03 & Scd & 114.6521 & 37.6335 & 431 & 51877 & 34 & 0.0130 & 0.8508 & 0.4333 & 0.4667 & 9.0732 & 0.6577 & 0.7980 \\
 UGC\_04226 & 1 & 0 & 1 & 0 & 6.624 & 4.436 & 3.355 & 5.3292 & 2.4379 & 8.9522 & 110.23 & Scd & 121.8392 & 40.3983 & 545 & 52202 & 111 & 0.0264 & 4.4377 & 0.0777 & 0.0161 &  & 4.4055 & 0.0253 \\
\enddata
\tablenotetext{a}{The full table is available in the electronic version of the paper.}
Columns:\\
$^{1}${Galaxy name.} \\
$^{2}${Whether galaxy belongs to the LOSS ``full-optimal'' subsample used to measure SN rates here and by L11 (1 = yes, 0 = no).} \\
$^{3\mbox{--}5}${Number of SNe~Ia, SE~SNe, and SNe~II (respectively) discovered in each galaxy.}\\
$^{6\mbox{--}8}${SN control time for SNe~Ia, SE~SNe, and SNe~II (respectively).}\\
$^{9}${Galaxy stellar mass, in units of $10^{10}~{\rm M_\sun}$, measured by L11 from the $B$- and $K$-band luminosities.}\\
$^{10\mbox{--}11}${$B$- and $K$-band luminosities (respectively), in units of $10^{10}~{\rm L_\sun}$.}\\
$^{12}${Distance to galaxy, in Mpc.}\\
$^{13}${Hubble type of galaxy, according to the system adopted by \citet{2011MNRAS.412.1419L}.}\\
$^{14\mbox{--}15}${Right ascension and declination (respectively) of galaxy, in decimal units.}\\
$^{16\mbox{--}18}${SDSS plate, Modified Julian Date (MJD), and Fiber identifier (respectively).}\\
$^{19}${Galaxy redshift, measured from the SDSS spectrum.}\\
$^{20\mbox{--}23}${Galaxy stellar mass, SFR, sSFR, and oxgen abundance (respectively) measured by the SDSS MPA-JHU Galspec pipeline, in units of $10^{10}~{\rm M_\sun}$, ${\rm M_\sun~yr^{-1}}$, $10^{-10}~{\rm yr^{-1}}$, and $12+{\rm log(O/H)}$ (respectively).}\\
$^{24\mbox{--}25}${Galaxy stellar mass and sSFR (respectively) measured from NSA Petrosian and Sersic photometry, in the same units as the MPA $M_\star$ and sSFR values.}\\
\end{splitdeluxetable*}

\newpage
\begin{sidewaystable*}
\vspace{-9.0cm}
\begin{deluxetable}{lCCCCCCCCCCCCC}
 \tablecaption{Mass-Normalized SN Rates as a Function of Galaxy Stellar Mass in  Different Hubble Types \label{table:rates}}
 \tablehead{
 \colhead{Galaxy type} & \colhead{$M_\star$\tablenotemark{a}} & \colhead{Rate\tablenotemark{b}} & \colhead{$N_{\rm SN}$} & \colhead{$M_\star$} & \colhead{Rate} & \colhead{$N_{\rm SN}$} & \colhead{$M_\star$} & \colhead{Rate} & \colhead{$N_{\rm SN}$} & \colhead{$M_\star$} & \colhead{Rate} & \colhead{$N_{\rm SN}$} & $N_{\rm tot}$ \\
 \colhead{} & \colhead{$(10^{10}~{\rm M}_\sun)$} & \colhead{(SNuM)} & \colhead{}  & \colhead{$(10^{10}~{\rm M}_\sun)$} & \colhead{(SNuM)} & \colhead{}  & \colhead{$(10^{10}~{\rm M}_\sun)$} & \colhead{(SNuM)} & \colhead{}  & \colhead{$(10^{10}~{\rm M}_\sun)$} & \colhead{(SNuM)} & \colhead{} & \colhead{}
 }
 \startdata
 \multicolumn{14}{c}{SNe Ia} \\
 E   & 6^{+4}_{-5}       & 0.09^{+0.04}_{-0.03} & 9  & 14.1^{+2.3}_{-1.6} & 0.10^{+0.05}_{-0.03}    & 8  & 21.1^{+2.6}_{-2.3} & 0.09^{+0.04}_{-0.03} & 9  & 33^{+15}_{-6} & 0.03^{+0.02}_{-0.01} & 9  & 35 \\
 S0  & 2.2^{+1.8}_{-1.7} & 0.18^{+0.06}_{-0.05} & 14 & 7.0^{+2.2}_{-1.7}  & 0.073^{+0.025}_{-0.019} & 14 & 14^{+4}_{-2}       & 0.052^{+0.018}_{-0.014} & 14 & 27^{+11}_{-5} & 0.051^{+0.018}_{-0.014} & 14 & 56 \\
 Sab & 1.6^{+1.2}_{-1.2} & 0.25^{+0.10}_{-0.08} & 11 & 4.4^{+0.5}_{-0.7}  & 0.19^{+0.08}_{-0.06}    & 10 & 7.5^{+2.4}_{-1.7}  & 0.07^{+0.03}_{-0.02} & 11 & 15.8^{+6.4}_{-3.5}  & 0.067^{+0.027}_{-0.020} & 11 & 43 \\
 Sb  & 1.2^{+1.0}_{-0.9}  & 0.28^{+0.11}_{-0.08} & 12 & 3.7^{+0.8}_{-0.7}  & 0.23^{+0.08}_{-0.06} & 13 & 7.4^{+2.6}_{-1.9}  & 0.067^{+0.025}_{-0.019} & 12 & 15.1^{+5.9}_{-2.7}  & 0.08^{+0.03}_{-0.02} & 13 & 50 \\
 Sbc & 1.0^{+1.1}_{-0.8} & 0.23^{+0.10}_{-0.08} & 9  & 3.9^{+0.8}_{-0.8}  & 0.17^{+0.08}_{-0.06} & 9  & 6.5^{+1.2}_{-1.0}  & 0.15^{+0.07}_{-0.05} & 9  & 11^{+6}_{-2}        & 0.078^{+0.035}_{-0.025} & 9  & 36 \\
 Sc  & 0.7^{+0.9}_{-0.5} & 0.24^{+0.12}_{-0.08} & 8  & 3.1^{+0.9}_{-0.7}  & 0.19^{+0.10}_{-0.07} & 8  & 6.0^{+1.2}_{-1.0}  & 0.18^{+0.09}_{-0.06} & 8  & 11.3^{+6.5}_{-2.7}  & 0.103^{+0.051}_{-0.036} & 8  & 32 \\
 Scd & 0.1^{+0.3}_{-0.1} & 0.22^{+0.21}_{-0.12} & 3  & 1.2^{+0.4}_{-0.3}  & 0.41^{+0.20}_{-0.14} & 8  & 2.9^{+1.3}_{-0.8}  & 0.11^{+0.08}_{-0.05} & 5  & 9.1^{+6.1}_{-2.5}   & 0.13^{+0.08}_{-0.05} & 6  & 22 \\
 E--Scd & 0.9^{+1.4}_{-0.8} & 0.24^{+0.03}_{-0.03} & 68 & 4.7^{+1.3}_{-1.0} & 0.140^{+0.019}_{-0.017} & 69 & 9.3^{+2.5}_{-1.9} & 0.086^{+0.012}_{-0.011} & 67 & 20^{+11}_{-5} & 0.056^{+0.008}_{-0.007} & 70 & 274 \\
 All galaxies & 0.8^{+1.4}_{-0.7} & 0.23^{+0.03}_{-0.03} & 68 & 4.6^{+1.3}_{-1.0} & 0.139^{+0.019}_{-0.017} & 69 & 9.3^{+2.5}_{-1.9} & 0.085^{+0.012}_{-0.010} & 67 & 20^{+11}_{-5} & 0.056^{+0.008}_{-0.007} & 70 & 274\\
 Passive\tablenotemark{c}      & 5.5^{+2.9}_{-3.4} & 0.069^{+0.037}_{-0.025} & 7  & 11.6^{+1.4}_{-1.0} & 0.12^{+0.06}_{-0.04} & 8  & 20^{+7}_{-4} & 0.043^{+0.023}_{-0.016} & 7  & & & & 22 \\ 
 Star forming\tablenotemark{c} & 0.3^{+0.7}_{-0.2} & 0.70^{+0.22}_{-0.17} & 16 & 2.7^{+1.1}_{-0.9}  & 0.22^{+0.07}_{-0.06} & 15 & 7.4^{+5.3}_{-2.3}  & 0.085^{+0.027}_{-0.021} & 16 & & & & 47 \\
 \multicolumn{14}{c}{SE SNe} \\ 
 E   & 1.2^{+1.6}_{-1.1} & 0.1^{+0.3}_{-0.1} & 1  & 8.8^{+2.8}_{-3.4} & 0.00^{+0.03}_{-0.00} & 0  & 22^{+11}_{-7} & 0.007^{+0.015}_{-0.006} & 1  & 56^{+21}_{-7} & 0.04^{+0.09}_{-0.03} & 1  & 3 \\
 S0  & 3^{+2}_{-2} & 0.010^{+0.024}_{-0.009} & 1  & 7.4^{+0.9}_{-0.9} & 0.018^{+0.042}_{-0.015} & 1  & 9.2^{+0.3}_{-0.3}   & 0.08^{+0.17}_{-0.06} & 1  & 15.8^{+10.8}_{-4.5} & 0.004^{+0.010}_{-0.004} & 1  & 4 \\  
 Sab & 0.8^{+0.8}_{-0.5} & 0.4^{+0.3}_{-0.2} & 4  & 2.7^{+0.5}_{-0.5} & 0.29^{+0.20}_{-0.13} & 5  & 4.5^{+0.6}_{-0.7}   & 0.12^{+0.10}_{-0.06} & 4  & 9^{+7}_{-3}   & 0.041^{+0.028}_{-0.018} & 5  & 18 \\
 Sb  & 0.6^{+0.5}_{-0.5} & 0.63^{+0.42}_{-0.27} & 5  & 1.9^{+0.2}_{-0.2} & 0.56^{+0.38}_{-0.24} & 5  & 4.1^{+1.5}_{-1.3}   & 0.11^{+0.07}_{-0.05} & 5  & 10^{+6}_{-3}  & 0.056^{+0.038}_{-0.024} & 5  & 20 \\
 Sbc & 0.7^{+0.7}_{-0.5} & 0.5^{+0.3}_{-0.2} & 5  & 3.3^{+1.4}_{-1.2} & 0.13^{+0.09}_{-0.06} & 5  & 6.6^{+0.6}_{-0.7}   & 0.36^{+0.21}_{-0.14} & 6  & 10.7^{+6.2}_{-2.5}  & 0.12^{+0.08}_{-0.05} & 5  & 21 \\
 Sc  & 0.5^{+0.6}_{-0.3} & 0.51^{+0.26}_{-0.19} & 7  & 2.3^{+0.9}_{-0.6} & 0.26^{+0.13}_{-0.09} & 8  & 4.8^{+1.2}_{-0.9}   & 0.24^{+0.13}_{-0.09} & 7  & 9^{+5}_{-2}   & 0.16^{+0.08}_{-0.06} & 8  & 30 \\
 Scd & 0.2^{+0.5}_{-0.2} & 0.23^{+0.18}_{-0.11} & 4  & 1.4^{+0.2}_{-0.2} & 0.54^{+0.43}_{-0.26} & 4  & 2.3^{+0.6}_{-0.5}   & 0.30^{+0.20}_{-0.13} & 5  & 5.2^{+5.7}_{-1.7}   & 0.09^{+0.07}_{-0.04} & 4  & 17 \\
 Irr & 0.1^{+0.1}_{-0.0} & 1.5^{+3.4}_{-1.2} & 1  & 0.2^{+0.0}_{-0.0} & 0^{+25}_{-0} & 0   & 0.4^{+1.1}_{-0.2}   & 0.4^{+0.8}_{-0.3} & 1  & 5.3^{+8.8}_{-2.4}   & 0.07^{+0.15}_{-0.06} & 1  & 3  \\
 Sab--Scd & 0.4^{+0.7}_{-0.3} & 0.40^{+0.10}_{-0.08} & 25 & 2.3^{+0.7}_{-0.6} & 0.26^{+0.06}_{-0.05} & 28 & 4.7^{+1.1}_{-0.9} & 0.15^{+0.04}_{-0.03} & 25 & 10.2^{+6.3}_{-2.7} & 0.088^{+0.020}_{-0.017} & 28 & 106 \\
 All galaxies & 0.4^{+0.7}_{-0.3} & 0.38^{+0.09}_{-0.07} & 28 & 2.4^{+0.7}_{-0.6} & 0.20^{+0.04}_{-0.04} & 30 & 4.9^{+1.2}_{-0.9} & 0.112^{+0.025}_{-0.021} & 29 & 12^{+10}_{-4} & 0.035^{+0.008}_{-0.007} & 29 & 116 \\
 \multicolumn{14}{c}{SNe II} \\
 S0  & 1.8^{+1.4}_{-1.4} & 0.016^{+0.037}_{-0.013} & 1  & 4.00^{+0.02}_{-0.03}  & 1.0^{+1.3}_{-0.6} & 2  & 5.2^{+1.0}_{-0.8} & 0.020^{+0.026}_{-0.013} & 2  & 12.9^{+10.8}_{-4.5} & 0.004^{+0.005}_{-0.002} & 2  & 7  \\
 Sab & 0.9^{+0.8}_{-0.6} & 0.65^{+0.26}_{-0.19} & 11 & 4^{+1}_{-1} & 0.13^{+0.05}_{-0.04} & 10 & 7.3^{+1.9}_{-1.4} & 0.10^{+0.04}_{-0.03} & 11 & 14^{+7}_{-3}  & 0.08^{+0.03}_{-0.02} & 11 & 43 \\
 Sb  & 1.4^{+1.2}_{-1.1} & 0.30^{+0.11}_{-0.08} & 14 & 4.4^{+0.9}_{-0.8} & 0.22^{+0.08}_{-0.06} & 13 & 8.0^{+1.7}_{-1.7} & 0.15^{+0.05}_{-0.04} & 14 & 14^{+6}_{-2}  & 0.12^{+0.04}_{-0.03} & 14 & 55 \\
 Sbc & 0.7^{+0.6}_{-0.5} & 1.3^{+0.4}_{-0.3} & 20 & 2.7^{+0.9}_{-0.7} & 0.45^{+0.13}_{-0.10} & 19 & 5.1^{+1.2}_{-0.8} & 0.38^{+0.11}_{-0.09} & 20 & 10.2^{+5.7}_{-2.5}  & 0.204^{+0.057}_{-0.045} & 20 & 79 \\
 Sc  & 0.4^{+0.5}_{-0.3} & 2.1^{+0.6}_{-0.5} & 17 & 1.5^{+0.3}_{-0.3} & 1.2^{+0.4}_{-0.3} & 17 & 3.5^{+1.8}_{-1.2} & 0.42^{+0.12}_{-0.10} & 18 & 9^{+6}_{-2}   & 0.40^{+0.12}_{-0.10} & 17 & 69 \\
 Scd & 0.1^{+0.2}_{-0.1} & 2.3^{+0.7}_{-0.6} & 16 & 0.7^{+0.3}_{-0.2} & 1.44^{+0.46}_{-0.36} & 16 & 1.7^{+0.5}_{-0.3} & 1.2^{+0.4}_{-0.3} & 17 & 4.1^{+5.0}_{-1.2}   & 0.36^{+0.11}_{-0.09} & 16 & 65 \\
 Irr & 0.05^{+0.04}_{-0.04} & 1.4^{+3.3}_{-1.2} & 1  & 0.4^{+1.1}_{-0.2} & 0.7^{+1.0}_{-0.5} & 2  & 3.3^{+0.8}_{-1.0} & 0.32^{+0.73}_{-0.26} & 1  & 9^{+7}_{-2}   & 0.19^{+0.26}_{-0.13} & 2  & 6  \\
 Sab--Scd & 0.3^{+0.5}_{-0.2} & 1.52^{+0.19}_{-0.17} & 78 & 2.0^{+0.7}_{-0.6} & 0.55^{+0.07}_{-0.06} & 77 & 4.6^{+1.4}_{-1.1} & 0.25^{+0.03}_{-0.03} & 78 & 11^{+6}_{-3} & 0.139^{+0.018}_{-0.016} & 78 & 311 \\
 All galaxies & 0.3^{+0.5}_{-0.2} & 1.33^{+0.16}_{-0.15} & 81 & 2.0^{+0.7}_{-0.6} & 0.43^{+0.05}_{-0.05} & 81 & 4.7^{+1.4}_{-1.1} & 0.18^{+0.02}_{-0.02} & 81 & 12^{+10}_{-4} & 0.052^{+0.006}_{-0.006} & 81 & 324 \\
 \enddata
 \tablenotetext{a}{The uncertainties mark the 16th and 84th percentiles of the mass distribution in a given bin.}
 \tablenotetext{b}{Mass-normalized SN rate, in units of $(10^{-12}~{\rm M}_\sun^{-1}~yr^{-1})$.}
 \tablenotetext{c}{Passive and star-forming galaxies are defined as having ${\rm log(sSFR/yr^{-1})}<-12$ and ${\rm log(sSFR/yr^{-1})}\geq-12$ (respectively), as measured by the MPA-JHU Galspec pipeline.}
\end{deluxetable}
\end{sidewaystable*}

\floattable
\begin{deluxetable*}{CCCC|CCCC|CCCC}
 \tablecaption{Specific SN Rates as a Function of Galaxy Stellar Mass, Measured with a Sliding Bin.\tablenotemark{a} \label{table:rate_snakes}}
 \tablehead{
 \colhead{$M_\star$} & \colhead{$R_{\rm Ia}$} & \colhead{$N_{\rm Ia}$} & \colhead{$N_{\rm gal}$} &
 \colhead{$M_\star$} & \colhead{$R_{\rm SE}$} & \colhead{$N_{\rm SE}$} & \colhead{$N_{\rm gal}$} &
 \colhead{$M_\star$} & \colhead{$R_{\rm II}$} & \colhead{$N_{\rm II}$} & \colhead{$N_{\rm gal}$} \\
 \colhead{$(10^{10}~{\rm M}_\sun)$} & \colhead{(SNuM)} & \colhead{} & \colhead{} &
 \colhead{$(10^{10}~{\rm M}_\sun)$} & \colhead{(SNuM)} & \colhead{} & \colhead{} &
 \colhead{$(10^{10}~{\rm M}_\sun)$} & \colhead{(SNuM)} & \colhead{} & \colhead{}
 }
 \colnumbers
 \startdata
 \multicolumn{4}{c}{SNe Ia} & \multicolumn{4}{c}{SE SNe} & \multicolumn{4}{c}{SNe II} \\
 \multicolumn{12}{c}{All galaxies} \\
 0.035^{+0.016}_{-0.013} & 2.2^{+5.1}_{-1.8} & 1 & 171 & 0.011^{+0.007}_{-0.007} & 8^{+19}_{-7}       & 1 & 166 & 0.015^{+0.011}_{-0.007} & 4.2^{+9.8}_{-3.5} & 1  & 189 \\
 0.084^{+0.036}_{-0.030} & 0.4^{+0.3}_{-0.9} & 1 & 454 & 0.075^{+0.045}_{-0.049} & 0.4^{+0.9}_{-0.3}  & 1 & 592 & 0.09^{+0.07}_{-0.04}    & 0.3^{+0.6}_{-0.2} & 1  & 623 \\
 0.09^{+0.04}_{-0.03}    & 1.1^{+1.1}_{-0.6} & 3 & 456 & 0.08^{+0.07}_{-0.05}    & 0.9^{+0.8}_{-0.5}  & 3 & 670 & 0.14^{+0.11}_{-0.06}    & 1.2^{+0.6}_{-0.4} & 8  & 728 \\
 0.10^{+0.06}_{-0.03}    & 0.9^{+0.8}_{-0.5} & 3 & 483 & 0.10^{+0.10}_{-0.05}    & 0.6^{+0.6}_{-0.3}  & 3 & 784 & 0.15^{+0.11}_{-0.07}    & 2.5^{+0.8}_{-0.6} & 17 & 727 \\
 0.14^{+0.07}_{-0.05}    & 0.9^{+0.7}_{-0.4} & 4 & 478 & 0.11^{+0.11}_{-0.06}    & 0.7^{+0.5}_{-0.3}  & 4 & 865 & 0.15^{+0.11}_{-0.07}    & 2.4^{+0.8}_{-0.6} & 16 & 713 \\
 \cdots & \cdots & \cdots & \cdots & \cdots & \cdots & \cdots & \cdots & \cdots & \cdots & \cdots & \cdots \\
 \cdots & \cdots & \cdots & \cdots & \cdots & \cdots & \cdots & \cdots & \cdots & \cdots & \cdots & \cdots \\
 \multicolumn{12}{c}{Specific Hubble types: E--Scd for SNe Ia, Sab--Scd for SE SNe and SNe II} \\
 0.04^{+0.05}_{-0.02}    & 3^{+6}_{-2}       & 1 & 143 & 0.012^{+0.018}_{-0.006} & 14^{+32}_{-11}       & 1 & 90  & 0.02^{+0.03}_{-0.01} & 7^{+16}_{-6} & 1 & 108 \\
 0.09^{+0.12}_{-0.05}    & 0.5^{+1.1}_{-0.4} & 1 & 374 & 0.050^{+0.071}_{-0.026} & 1.6^{+3.8}_{-1.4}    & 1 & 208 & 0.09^{+0.13}_{-0.05} & 0.5^{+1.2}_{-0.4}  & 1 & 372 \\
 0.09^{+0.13}_{-0.06}    & 1.3^{+1.3}_{-0.7} & 3 & 379 & 0.08^{+0.12}_{-0.04}    & 1.2^{+1.5}_{-0.7}    & 2 & 373 & 0.10^{+0.15}_{-0.06} & 2.3^{+1.4}_{-0.9}  & 6 & 402 \\
 0.10^{+0.16}_{-0.07}    & 1.0^{+1.0}_{-0.5} & 3 & 416 & 0.2^{+0.3}_{-0.1}       & 0.3^{+0.4}_{-0.2}    & 2 & 601 & 0.13^{+0.20}_{-0.08} & 1.7^{+1.0}_{-0.7}  & 6 & 435 \\
 0.14^{+0.21}_{-0.09}    & 1.0^{+0.8}_{-0.5} & 4 & 430 & 0.3^{+0.5}_{-0.2}        & 0.2^{+0.2}_{-0.1}    & 3 & 744 & 0.13^{+0.21}_{-0.08} & 2.3^{+1.1}_{-0.8}  & 8 & 433 \\
 \cdots & \cdots & \cdots & \cdots & \cdots & \cdots & \cdots & \cdots & \cdots & \cdots & \cdots & \cdots \\
 \cdots & \cdots & \cdots & \cdots & \cdots & \cdots & \cdots & \cdots & \cdots & \cdots & \cdots & \cdots \\
 \enddata
\tablenotetext{a}{The electronic version of the paper contains six separate versions of this table, one for each SN type.}
Columns:\\
$^{1,5,9}${Stellar mass of the galaxies in the bin. Error bars represent the upper and lower limits of the bin.}\\
$^{2,6,10}${Specific SN rate. Error bars are 68\% Poisson uncertainties stemming from the number of SNe in each bin.}\\
$^{3,7,11}${Number of SNe in the given bin.}\\
$^{4,8,12}${Number of galaxies in the given bin.}\\
\end{deluxetable*}

\floattable
\begin{deluxetable}{CCCCC}
 \tablecaption{Ratio Between SE SN and SN II Specific Rates in the Range $2\times10^{9}\leq M_\star \leq 2\times10^{11}~{\rm M}_\sun$. \label{table:rate_ratio}}
 \tablehead{
 \colhead{Mass} & \colhead{Metallicity} & \colhead{$R_{\rm SE}/R_{\rm II}$} & \colhead{$N_{\rm SE}$} & \colhead{$N_{\rm II}$} \\
 \colhead{($10^{10}~{\rm M}_\sun$)} & \colhead{$12+{\rm log(O/H)}$} & \colhead{} & \colhead{} & \colhead{}
 }
 \colnumbers
 \startdata
 0.32^{+0.12}_{-0.09} & 8.84^{+0.06}_{-0.06} & 0.13^{+0.09}_{-0.08} & 3  & 26 \\
 0.84^{+0.28}_{-0.23} & 9.01^{+0.05}_{-0.06} & 0.35^{+0.11}_{-0.11} & 14 & 47 \\
 2.1^{+0.7}_{-0.6}    & 9.17^{+0.05}_{-0.05} & 0.59^{+0.13}_{-0.11} & 34 & 77 \\
 5.1^{+1.8}_{-1.3}    & 9.32^{+0.05}_{-0.05} & 0.64^{+0.14}_{-0.11} & 38 & 93 \\
 12^{+4}_{-3}         & 9.47^{+0.05}_{-0.05} & 0.64^{+0.15}_{-0.15} & 20 & 58 \\
 \enddata
 \tablenotetext{1}{The uncertainties mark the 16th and 84th percentiles of the mass distribution in a given bin.}
 \tablenotetext{2}{Metallicity (on the T04 scale) converted from stellar masses using the mass-metallicity relation in Table~\ref{table:fits} (for all galaxies).}
 \tablenotetext{3}{Ratio between mass-normalized SE SN and SN II rates.}
 \tablenotetext{4}{Number of SE SNe in a given bin.}
 \tablenotetext{5}{Number of SNe II in a given bin.}
\end{deluxetable}

\floattable
\begin{deluxetable*}{cCCCCCCCCC}
 \tablecaption{Specific SN Rates as a Function of Various Galaxy Properties. \label{table:rates_OH}}
 \tablehead{\colhead{SN type} & \colhead{Metallicity\tablenotemark{a}} & \colhead{$R$} & \colhead{$N_{\rm SN}$} & \colhead{sSFR} & \colhead{$R$} & \colhead{$N_{\rm SN}$} & \colhead{SFR} & \colhead{$R$} & \colhead{$N_{\rm SN}$} \\
 \colhead{} & \colhead{$12+{\rm log(O/H)}$} & \colhead{(SNuM)} & \colhead{} & \colhead{${\rm log(sSFR/yr^{-1})}$} & \colhead{(SNuM)} & \colhead{} & \colhead{${\rm log(SFR}/{\rm M}_\sun~{\rm yr^{-1})}$} & \colhead{(SNuM)} & \colhead{}}
 \colnumbers
 \startdata
  \multirow{4}{*}{Ia}   & 8.81^{+0.18}_{-0.25}    & 0.77^{+0.38}_{-0.27} & 8  & -12.16^{+0.19}_{-0.25} & 0.09^{+0.03}_{-0.02}    & 16 & -1.2^{+0.3}_{-0.4}    & 0.118^{+0.032}_{-0.026} & 21 \\
                        & 9.109^{+0.035}_{-0.052} & 0.20^{+0.10}_{-0.07} & 8  & -11.2^{+0.5}_{-0.5}    & 0.073^{+0.023}_{-0.018} & 16 & -0.23^{+0.27}_{-0.34} & 0.100^{+0.026}_{-0.021} & 22 \\
                        & 9.19^{+0.08}_{-0.02}    & 0.31^{+0.15}_{-0.11} & 8  & -10.33^{+0.12}_{-0.14} & 0.16^{+0.05}_{-0.04}    & 17 &  0.34^{+0.25}_{-0.14} & 0.135^{+0.035}_{-0.029} & 22 \\
                        &                         &                      &    & -9.80^{+0.38}_{-0.25}  & 0.31^{+0.10}_{-0.08}    & 16 & \\           
  \hline
  \multirow{3}{*}{SE}   & 8.77^{+0.18}_{-0.23}    & 1.0^{+0.7}_{-0.4}    & 5  & -11.84^{+0.57}_{-0.45} & 0.008^{+0.019}_{-0.007} & 1 & -0.68^{+0.45}_{-0.64}  & 0.045^{+0.027}_{-0.018} & 6  \\
                        & 9.07^{+0.04}_{-0.04}    & 0.34^{+0.27}_{-0.16} & 4  & -10.43^{+0.17}_{-0.30} & 0.09^{+0.03}_{-0.05}    & 6 & 0.11^{+0.13}_{-0.13}   & 0.13^{+0.08}_{-0.05}    & 6  \\ 
                        & 9.166^{+0.072}_{-0.035} & 0.25^{+0.17}_{-0.11} & 5  & -10.08^{+0.09}_{-0.07} & 0.28^{+0.19}_{-0.12}    & 5 & 0.45^{+0.21}_{-0.12}   & 0.13^{+0.08}_{-0.05}    & 6  \\
                        &                         &                      &    & -9.7^{+0.3}_{-0.2}     & 0.35^{+0.21}_{-0.14}    & 6 & \\
  \hline
  \multirow{4}{*}{II}   & 8.71^{+0.16}_{-0.25}    & 3.4^{+1.3}_{-1.0}    & 12 & -11.84^{+0.57}_{-0.45} & 0.033^{+0.015}_{-0.011} & 9  & -0.8^{+0.4}_{-0.6}    & 0.08^{+0.02}_{-0.02}    & 20 \\
                        & 9.01^{+0.05}_{-0.04}    & 1.37^{+0.47}_{-0.36} & 14 & -11.51^{+0.15}_{-0.28} & 0.17^{+0.05}_{-0.04}    & 17 & -0.1^{+0.1}_{-0.1}    & 0.30^{+0.09}_{-0.07}    & 19 \\
                        & 9.112^{+0.018}_{-0.022} & 0.73^{+0.26}_{-0.20} & 13 & -10.20^{+0.06}_{-0.08} & 0.56^{+0.17}_{-0.13}    & 18 & 0.22^{+0.09}_{-0.08}  & 0.35^{+0.10}_{-0.08}    & 20 \\
                        & 9.18^{+0.07}_{-0.03}    & 0.49^{+0.18}_{-0.13} & 13 & -10.0^{+0.1}_{-0.1}    & 0.9^{+0.3}_{-0.2}       & 17 & 0.5^{+0.2}_{-0.1}     & 0.32^{+0.09}_{-0.07}    & 20 \\
                        &                         &                      &    & -9.55^{+0.34}_{-0.17}  & 1.4^{+0.4}_{-0.3}       & 18 & \\
 \enddata
 \tablenotetext{a}{SDSS metallicity values were measured using the T04 metallicity scale.}
\end{deluxetable*}

\floattable
\begin{deluxetable}{CCCCCC}
 \tablecaption{Polynomial Fits ($y=ax+b$) to SN Rates and Galaxy Scaling Relations. \label{table:fits}}
 \tablehead{\colhead{$y$} & \colhead{$x$} & \colhead{$a$} & \colhead{$b$} & \colhead{$\chi^2/$DOF} & \colhead{$S$}}
 \colnumbers
 \startdata
 {\rm log}(R_{\rm Ia}/{\rm SNuM})\tablenotemark{a} & {\rm log}(M_\star/{\rm M}_\sun) & -0.44 \pm 0.06 & 3.8 \pm 0.6 & 2.7/2     & >5\sigma \\
 {\rm log}(R_{\rm Ia}/{\rm SNuM})\tablenotemark{b} & {\rm log}(M_\star/{\rm M}_\sun) & -0.43 \pm 0.08 & 3.7 \pm 0.8 & 3.9/2     & >5\sigma \\
 {\rm log}(R_{\rm Ia}/{\rm SNuM})                  & {\rm log(sSFR/yr^{-1})}         & 0.23 \pm 0.08  & 1.7 \pm 0.8 & 3.2/2     & >3\sigma \\
 {\rm log}(R_{\rm Ia}/{\rm SNuM})                  & 12+{\rm log(O/H)}               & -1.3 \pm 0.7   & 11 \pm 7    & 1.0/1     & >2\sigma \\
 \hline
 {\rm log}(R_{\rm SE}/{\rm SNuM})\tablenotemark{a} & {\rm log}(M_\star/{\rm M}_\sun) & -0.64 \pm 0.09 & 5.8 \pm 0.9 & 7.3/2     & >4\sigma \\
 {\rm log}(R_{\rm SE}/{\rm SNuM})\tablenotemark{b} & {\rm log}(M_\star/{\rm M}_\sun) & -0.46 \pm 0.10 & 4.1 \pm 1.0 & 2.1/2     & >4\sigma \\
 {\rm log}(R_{\rm SE}/{\rm SNuM})\tablenotemark{c} & {\rm log(sSFR/yr^{-1})}         & 0.8 \pm 0.5    & 7 \pm 5     & 0.5/1     & >5\sigma \\
 {\rm log}(R_{\rm SE}/{\rm SNuM})                  & 12+{\rm log(O/H)}               & -1.5 \pm 1.0   & 13 \pm 9    & 10^{-5}/1 & >2\sigma \\
 \hline
 {\rm log}(R_{\rm II}/{\rm SNuM})\tablenotemark{a}   & {\rm log}(M_\star/{\rm M}_\sun) & -0.84 \pm 0.05 & 8.2 \pm 0.5   & 21/4  & >4\sigma \\
 {\rm log}(R_{\rm II}/{\rm SNuM})\tablenotemark{b}   & {\rm log}(M_\star/{\rm M}_\sun) & -0.68 \pm 0.05 & 6.6 \pm 0.5   & 5.6/4  & >4\sigma \\
 {\rm log}(R_{\rm II}/{\rm SNuM})\tablenotemark{c}   & {\rm log(sSFR/yr^{-1})}         & 0.9 \pm 0.2    & 9 \pm 2       & 2.7/2  & >5\sigma \\
 {\rm log}(R_{\rm II}/{\rm SNuM})                    & 12+{\rm log(O/H)}               & -1.73 \pm 0.45 & 16 \pm 4      & 0.6/2  & >5\sigma \\
 \hline
 {\rm log}(M_{\star})\tablenotemark{d}               & 12+{\rm log(O/H)}         & 2.5              & -12.6             &           & \\
 {\rm log}(M_{\star,\rm Ia})                         & 12+{\rm log(O/H)}         & 2.31 \pm 0.13    & -10.8 \pm 1.2     & 481/24    & \\
 {\rm log}(M_{\star,\rm SE})                         & 12+{\rm log(O/H)}         & 3.4 \pm 0.1      & -20.4 \pm 1.2     & 16/10     & \\
 {\rm log}(M_{\star,\rm II})                         & 12+{\rm log(O/H)}         & 3.33 \pm 0.08    & -20.1 \pm 0.7     & 321/45    & \\
 \enddata
 \tablenotetext{a}{Using rates measured for all galaxy types.}
 \tablenotetext{b}{Using rates measured in E--Scd (Sab--Scd) galaxies for SNe Ia (SE SNe and SNe II).}
 \tablenotetext{c}{Restricted to measurements in galaxies with ${\rm log(sSFR/yr^{-1})}>-11$.}
 \tablenotetext{d}{Measured with MATLAB's cftool fitting suite, which does not provide $\chi^2$ values.}
\end{deluxetable}

\end{document}